\documentclass[12pt,preprint]{aastex}

\newcommand  \cmsq     {\ifmmode {\rm cm}^{-2} \else cm$^{-2}$\fi}
\newcommand  \ergs     {\ifmmode {\rm erg\,s}^{-1} \else erg s$^{-1}$\fi}
\newcommand  \ergcms   {\ifmmode {\rm erg\,cm}^{-2}\,{\rm s}^{-1}
                        \else erg\,cm$^{-2}$\,s$^{-1}$\fi}
\def \Msun {\ifmmode M_{\odot} \else $M_{\odot}$\fi}
\def \Lsun {\ifmmode L_{\odot} \else $L_{\odot}$\fi}
\def \spitzer  {{\it Spitzer}}
\def \chandra  {{\it Chandra}}
\def \rosat    {{\it ROSAT}}
\def \xmm      {{\it XMM}}
\def \herschel {{\it Herschel}}
\def \hst      {{\it HST}}
\def \swift    {{\it Swift}}
\def \iras     {{\it IRAS}}

\slugcomment{Accepted for publication in ApJ}

\shorttitle{Star formation in (E)CDFS AGN hosts}
\shortauthors{Lutz et al.}

\begin{document}

\title{The LABOCA survey of the Extended Chandra Deep Field South: Two 
modes of star formation in AGN hosts?}

\author{D. Lutz\altaffilmark{1},
V. Mainieri\altaffilmark{2},
D. Rafferty\altaffilmark{3},
L. Shao\altaffilmark{1},
G. Hasinger\altaffilmark{4},
A. Wei\ss\altaffilmark{5},
F. Walter\altaffilmark{6},
I. Smail\altaffilmark{7},
D.M. Alexander\altaffilmark{8},
W.N. Brandt\altaffilmark{3},
S. Chapman\altaffilmark{9},
K. Coppin\altaffilmark{7},
N.M. F\"orster Schreiber\altaffilmark{1},
E. Gawiser\altaffilmark{10},
R. Genzel\altaffilmark{1},
T.R. Greve\altaffilmark{6},
R.J. Ivison\altaffilmark{11,12},
A.M. Koekemoer\altaffilmark{13},
P. Kurczynski\altaffilmark{10},
K.M. Menten\altaffilmark{5},
R. Nordon\altaffilmark{1},
P. Popesso\altaffilmark{1},
E. Schinnerer\altaffilmark{6},
J.D. Silverman\altaffilmark{14},
J. Wardlow\altaffilmark{8},
Y.Q. Xue\altaffilmark{3}
\altaffiltext{1}{Max-Planck-Institut f\"ur extraterrestrische Physik,
Postfach 1312, 85741 Garching, Germany \email{lutz@mpe.mpg.de}}
\altaffiltext{2}{European Southern Observatory, Karl-Schwarzschild-Stra\ss e 2,
85748 Garching, Germany}
\altaffiltext{3}{Department of Astronomy and Astrophysics, 525 Davey Lab,
Pennsylvania State University, University Park, PA 16802, USA} 
\altaffiltext{4}{Max-Planck-Institut f\"ur Plasmaphysik, Boltzmannstra\ss e 2,
85748 Garching, Germany}
\altaffiltext{5}{Max-Planck-Institut f\"ur Radiastronomie, Auf dem H\"ugel 69,
53121 Bonn, Germany}
\altaffiltext{6}{Max-Planck-Institut f\"ur Astronomie, K\"onigstuhl 17, 69117
Heidelberg, Germany}
\altaffiltext{7}{Institute for Computational Cosmology, Durham University,
South Road, Durham, DH1 3LE, UK}
\altaffiltext{8}{Department of Physics, Durham University, South Road,
Durham, DH1 3LE, UK}
\altaffiltext{9}{Institute of Astronomy, Madingley Road, Cambridge, CB3 0HA, U.K.}
\altaffiltext{10}{Physics \& Astronomy Department, Rutgers University,
Piscataway, NJ 08854, USA}
\altaffiltext{11}{SUPA, Institute for Astronomy, University of Edinburgh,
Royal Observatory, Blackford Hill, Edinburgh, EH9 3HJ, UK}
\altaffiltext{12}{UK Astronomy Technology Centre,
Royal Observatory, Blackford Hill, Edinburgh, EH9 3HJ, UK}
\altaffiltext{13}{Space Telescope Science Institute, 3700 San Martin Drive,
Baltimore, MD 21218, USA}
\altaffiltext{14}{Institute for the Physics and Mathematics of the  
Universe (IPMU), University of Tokyo, Kashiwanoha 5-1-5, Kashiwa,  
Chiba 277-8568, Japan}
}

\begin{abstract}
We study the co-existence of star formation and AGN activity in \chandra\ 
X-ray selected AGN by analyzing stacked 870$\mu$m submm emission 
from a deep and wide
map of the Extended Chandra Deep Field South, obtained with the LABOCA 
instrument at the APEX telescope. The total X-ray sample of 895 sources
with median redshift z$\sim$1 
drawn from the combined (E)CDFS X-ray catalogs is detected 
at $>11\sigma$  significance at a mean submm flux of $0.49\pm 0.04$mJy, 
corresponding
to a typical star formation rate around 30\Msun yr$^{-1}$ for a 
T=35K, $\beta$=1.5 greybody far-infrared spectral energy distribution. 
The good signal to 
noise ratio permits stacking analyses for major subgroups, splitting the
sample by redshift, intrinsic luminosity, and AGN obscuration properties. We 
observe a trend of star formation rate increasing with redshift. An 
increase of star formation rate with AGN luminosity is indicated at the 
highest $L_{2-10keV}\gtrsim 10^{44}\ergs$ luminosities only.
Increasing trends with X-ray obscuration as expected in some AGN 
evolutionary scenarios are not observed for the bulk of the X-ray AGN sample
but may be present for the highest intrinsic luminosity objects with 
$L_{2-10keV}\gtrsim 10^{44}\ergs$. This 
behaviour suggests a transition between two modes in the coexistence of 
AGN activity and star formation. For the bulk of the sample,  
the X-ray luminosity and obscuration of
the AGN are not intimately linked to the global star formation rate of their 
hosts. The hosts are likely massive and
forming stars secularly, at rates similar to the pervasive star formation 
seen in massive galaxies without an AGN at similar redshifts. In these 
systems, star formation is not linked to a specific state of the AGN and 
the period of moderately luminous  AGN activity may not highlight a 
major evolutionary transition of the galaxy. 
The change indicated towards more intense star formation, and a more 
pronounced increase in star formation rates between unobscured and 
obscured AGN reported in the literature at highest 
($L_{2-10keV}\gtrsim 10^{44}\ergs$) luminosities suggests 
that these luminous AGN follow an evolutionary path on which 
obscured AGN activity and intense star formation are linked, possibly 
via merging. Comparison to local hard X-ray selected AGN
supports this interpretation. Star formation rates
in the hosts of moderate luminosity AGN at z$\sim$1 are an order of magnitude 
higher than at z$\sim$0, following the increase in the non-AGN massive galaxy 
population. At high AGN luminosities, hosts on the evolutionary link/merger 
path emerge from this secular level of star formation.
 
\end{abstract}


\section{Introduction}

Observing the co-evolution of active galactic nuclei (AGN) and
their hosts is key to understanding the similar 
cosmic evolution of the space density of luminous AGN and of the
star formation rate density. This co-evolution also has to lead to today's 
fossil relations between
remnant supermassive black hole mass and the properties of the host spheroid. 
Directly measuring the star formation rate of a high redshift galaxy is 
however  particularly difficult for AGN hosts, since the AGN proper will 
disturb and overwhelm the rest frame UV and optical spectral and 
photometric star formation tracers already at modest 
luminosities,
unless the AGN is obscured. Furthermore, star formation in the host may be 
noticeably obscured in particular at high star formation rates approaching
that of high-redshift ultraluminous infrared and submillimeter galaxies. 
Infrared observations can thus play an important role for these studies. One 
approach that has been
successfully applied to high redshift AGN is to use the mid-infrared 
polycyclic aromatic hydrocarbon (`PAH')
emission features which can be separated from the strong AGN mid-IR
continuum emission by means of low resolution spectroscopy 
\citep[e.g.,][]{houck05,lutz05}. However, even with the superb
spectroscopic sensitivity of the \spitzer\ Space Telescope, this approach 
is limited for high redshifts to modest sample sizes. 
Alternatively, observations of
the rest frame far-infrared/submm continuum have been used since the advent
of the first sensitive submm photometers to measure star formation via the 
far-infrared/submm continuum from the associated cool (T$\sim$35K) dust. 
This rests on the submm continuum being due to star 
formation in the host galaxy, with star formation dominating over the AGN 
heated dust emission at these wavelengths for all but 
the highest ratios of AGN luminosity to star formation in the host. Such a 
star formation dominance in the submm is 
possible because of the steep decline of AGN dust emission towards 
far-infrared and submm wavelengths 
\citep[e.g.,][and later torus models]{pier92}, while
the SEDs of star forming galaxies have a pronounced far-infrared (FIR) peak. 
The assumption of star formation dominating the submm emission
is supported by several types of observations. For NGC 1068, the 
AGN in the local universe for which current spatial resolution is sufficent to 
separate the 
AGN from the host in the far-infrared/submillimeter range, this assumption is 
directly supported by observations \citep{pier93,papadopoulos99,lefloch01}.  
Note that NGC 1068 falls well into the range of AGN luminosities 
\citep[intrinsic $L_{2-10keV}\sim 10^{43.5}\,\ergs$,]
[]{colbert02} and far-infrared luminosities ($L_{IR}\sim 10^{11.3}\,\Lsun$) of 
the AGN found in deep X-ray surveys, such as the (E)CDFS AGN that are 
discussed below.
Recent \spitzer\ spectroscopic studies using the polycyclic aromatic 
hydrocarbon (PAH) emission features in comparison to the far-infrared/submm 
emission lend further support to this assumption, indicating that,
at $L_{AGN}/L_{FIR}\sim 10$, the far-infrared emission
of local QSOs as well as of mm-bright high redshift QSOs is
dominated by star formation in the host \citep{schweitzer06,netzer07,lutz08},
plausibly fed by large reservoirs of molecular gas 
\citep[e.g.,][]{evans01,solomon05}. 
Here, we study a sample that is at or below this ratio 
of AGN and star forming luminosities
(\S3.2). Our sample  does not reach the extreme values as some of the most 
luminous
but mm-faint high-z QSOs, for which the assignment of far-infrared emission to
star formation is more uncertain \citep[see discussion in][]{lutz08}.  
Our sample is X-ray selected and thus not biased towards extremely 
obscured infrared
objects as, e.g., some local ULIRGs or high redshift dust-obscured galaxies 
\citep[e.g.,][]{houck05}, for which the AGN contribution to the far-infrared
emission may be significant.
 
Measuring the submm emission and on its basis the star formation rate of 
AGN hosts thus plays an important role in constraining the evolution of AGNs. 
Such studies directly benefit from the improving depth and 
areal coverage of current submm surveys. Both, the study of local AGN and 
of the AGN/galaxy (co)evolution 
have lead to the suggestion of models in which intense star formation events
and powerful AGN activity are physically linked and sequentially occur in a
single object \citep[e.g.,][]{sanders88,fabian99,granato04,hopkins06}, in a 
process closely linked to the hierarchical 
merging of galaxies in the universe. In a nutshell, galaxy interaction followed
by merging with associated gas inflow may trigger a powerful burst of star 
formation 
and subsequently feed the central black hole(s) of the merger, to produce a 
luminous AGN.
The AGN may then quench the star formation event and shed the obscuring dust, 
and finally emerge as an optically visible
QSO. In this evolutionary picture, an obscured AGN would sample the 
earlier phases of the AGN activity and would be associated with more 
powerful star formation in the AGN host, and thus stronger submm emission.
As already implied by the typically non-merger morphology of local 
moderate-luminosity AGN (Seyferts), such an evolutionary picture may not 
always apply and its applicability needs to be studied as a function of 
AGN luminosity and redshift.

A different behaviour would be expected in the context of the successful
unified picture of active galactic nuclei proper, in which the differences 
between obscured and unobscured
types of AGN are the consequences of different viewing directions on
intrinsically identically structured systems. In fact, for the AGN and its 
immediate dusty surroundings, for example in the form of a `torus', 
emission would be expected 
to be equally strong or stronger in the face-on/unobscured direction compared
to the edge-on direction, even at long far-infrared or submm 
wavelengths \citep[e.g.,][]{pier92,nenkova08}. However, because of the 
intrinsically weaker AGN/torus emission at far-infrared and submm wavelengths, 
such differences will be easily washed out by even a 
modest amount of star formation in the host galaxy. This would lead to the 
expectation
of little dependency of submm emission to obscuration in the unified AGN 
picture. This `unified' view is by no means contradictory to the 
evolutionary picture.
Considering both these perspectives only stresses that differences in submm 
emission found between 
different AGN types will to a large extent reflect
evolutionary coupling of periods of AGN activity and host star formation, 
rather than AGN physics and orientation alone. If evolutionary signatures 
are found in the submm, the underlying mechanisms like merging have to be 
clarified from 
additional information like morphology or dynamics. Combining evolutionary 
and unified perspectives 
also emphasizes the need to separately test for possible evolutionary 
effects in populations of high redshift AGN of different luminosities, that 
may follow different paths of AGN and host evolution.

First looks at the submm emission of X-ray selected AGN have 
compared deep X-ray surveys to the first 
generation of (sub)mm surveys \citep[e.g.,][]{fabian00, severgnini00,
hornschemeier00,barger01,almaini03,waskett03}. In 
general, no straightforward correspondence
between typical sources from these X-ray surveys and bright 
submm sources detected with SCUBA 
\citep{holland99} was 
established, and the average flux from submm stacking experiments of X-ray AGN
was found to be low, e.g.,
S$_{850\mu m}=1.21\pm 0.27$mJy for the Chandra Deep field North (CDFN)
\citep{barger01} and
S$_{850\mu m}=0.48\pm 0.27$mJy from the Canada-UK deep SCUBA survey (CUDSS) 
\citep{waskett03}, with insufficient S/N of the stacks
for an in-depth analysis as a function of AGN properties. Pointed 
submm followup of selected, e.g.,  hard/luminous sources from the deep 
X-ray surveys resulted in a few detections 
\citep[e.g.,][]{mainieri05,rigopoulou09} but also some upper limits. 
Conversely, SCUBA sources were often found to be associated with 
X-ray faint AGN in the very deepest X-ray surveys 
\citep{alexander05a,alexander05b}. These faint AGN do not dominate the 
energetics of the SCUBA sources and were undetectable in earlier analyses 
that were not based on the ultradeep 2Ms \chandra\ data. Their black hole 
masses appear modest compared to similarly massive galaxies and to more
powerful AGN \citep{borys05,alexander08}.
 
The study that has been perhaps most successful so far in establishing 
bright submm 
emission for an X-ray selected AGN population and finding trends with AGN 
properties used a different approach. Selecting 
luminous X-ray absorbed but optically bright AGN not from deep field \chandra\ 
or \xmm\ X-ray survey data but from 
identification of X-ray brighter sources from the \rosat\ survey,
 \citet{page01} were able to detect 4 of 8 
X-ray absorbed QSOs at S$_{850\mu m}>5$mJy. Later comparisons with matched 
X-ray unabsorbed samples \citep{page04,stevens05} provided evidence for a 
lower submm detection rate of the X-ray unabsorbed objects. This difference 
in submm brightness between obscured and unobscured objects
supports the evolutionary view, but questions remain in particular when
comparing the large submm fluxes for some of the \citet{page01} objects with
the more modest success of submm follow up observations of luminous hard 
sources from deep fields. If submm emission follows the emission at 
other wavelengths, this might simply 
reflect the brighter observed fluxes and larger AGN luminosities of the
\citet{page01} sources. However, it should be noted that these are selected as
moderately X-ray obscured ($N_H\sim 10^{22}\cmsq$) but optically 
unobscured (optical broad emission line = Type 1) AGN, unlike many of the 
heavily obscured AGN from deep X-ray fields, which show Type 2 optical 
spectra lacking broad lines. Like the broad lines, their bright optical 
magnitudes \citep{page01b} argue against a significant obscuration of the AGN
in the rest frame optical/UV. Extending such studies to include more typical 
obscured AGN is clearly important and a main motivation of this paper.

We here present a study of the submm properties of X-ray selected AGN
in the (extended) Chandra Deep Field South (E)CDFS. Making use of a new submm
map provided by LABOCA at the APEX facility \citep{weiss09} as well as 
current X-ray data with substantially improved identification 
status and characterisation of the AGN, we can study the submm properties 
and hence host star formation rates as a function of AGN properties. 
Throughout the paper, we adopt an $\Omega_m =0.3$, $\Omega_\Lambda =0.7$ and 
$H_0=70$ km\,s$^{-1}$\,Mpc$^{-1}$ cosmology. When distinguishing between
moderate luminosity (Seyfert) and high luminosity (QSO) AGN we refer to
an intrinsic luminosity limit of $L_{2-10keV}=10^{44}\ergs$ unless stated 
otherwise.

\section{Results from submillimeter mapping of the Extended Chandra Deep 
Field South}
The excellent X-ray and multiwavelength coverage of the 0.1 square degree 
Chandra Deep Field
South (CDFS) and the surrounding 0.3 square degree Extended Chandra Deep 
Field South (ECDFS), 
in combination 
with the powerful mapping capabilities of the LABOCA submm camera 
\citep{siringo09} at the APEX telescope \citep{guesten06}, enables us 
to take a fresh look 
at the issues discussed in the introduction, making use of the improved 
observational resources.  We use the
LABOCA 870$\mu$m map obtained by the LESS (LABOCA ECDFS Submm Survey) 
consortium as described by 
\citet{weiss09}. We use the beam-convolved v2.2 final map which includes a 
total of about 350 hours of observing time over an area of about 
$40\arcmin\times 40\arcmin$ with an rms noise level of 
about 1.2\,mJy\,beam$^{-1}$ in the inner $30\arcmin\times 30\arcmin$. 
This map and the catalog of 126 sources detected at $>3.7\sigma$ is
presented in \citet{weiss09}. We also make use of the residual 
map obtained by subtracting these 126 sources, using fluxes that consider
in a statistical sense the boosting by instrument and confusion noise 
\citep{weiss09,coppin05}. 

\subsection{Samples of X-ray AGN}

To maximise the statistics and to allow us to draw meaningful conclusions 
for physically
selected subgroups of X-ray selected AGN, we use X-ray based AGN samples 
from \chandra\ 
observations for both the ECDFS and the deeper but smaller CDFS, for
a total of 895 X-ray sources.
For the CDFS, we mainly use X-ray spectral properties
of \citet{tozzi06} which are related to and based on the original CDFS 
observations of \citet{giacconi02}. Recently, the deeper 2~Msec \chandra\ 
data and catalog for the CDFS have become available \citep{luo08} but 
physical modelling comparable to the level of \citet{tozzi06} is not yet 
completed (Bauer et al., in preparation). We therefore use the updated 
2~Msec observational data 
(observed fluxes, hardness ratios etc.), from \citet{luo08} where available
but stick to the \citet{tozzi06} results for physical properties derived
from X-ray spectral fitting (intrinsic luminosities, X-ray obscuring column 
density $N_H$, etc.). We have also added 94 CDFS sources that are new from
\citet{luo08}. Lacking X-ray spectral fitting, these were used only for the
combined stack and when analysing properties as a function of redshift, no 
modelled properties are available for those. 
To reduce contamination by non-AGN sources we exclude here objects 
that are likely nearby normal/star forming galaxies. Discrimination 
between such objects and AGN is approximately possible by comparing their 
X-ray to their
optical properties; specifically we have adopted a cutoff 
$log(f_X/f_R)\geq -1$ \citep[e.g.,][]{bauer04}, for the ratio of the X-ray 
flux to the
optical R-band flux. Here we use as X-ray flux the observed full
band flux if available, otherwise the larger of hard and soft band flux. 
For similar reasons, we exclude 14 sources from \citet{tozzi06} with 
intrinsic rest frame $L_{0.5-10keV}<10^{41}\ergs$. Sources with such
low X-ray luminosities will mostly not be AGN \citep[e.g.,][]{bauer04}. 
We have furthermore excluded 20 sources from 
\citet{tozzi06} that are not re-detected in the deeper 2Msec data of 
\citet{luo08}. In 
particular, this cut includes the \citet{giacconi02} XID 618. While this 
source, indicated in only one of two source extraction methods used by 
\citet{giacconi02}, is coincident with an 
interesting z=4.76 submillimeter galaxy (SMG) \citep{coppin09}, its nature 
as an X-ray source is 
not confirmed by the deeper X-ray data of \citet{luo08}. Including it in
our analysis as a luminous obscured X-ray source as inferred by 
\citet{tozzi06} would increase the differences between luminous unobscured
and obscured AGN that we discuss in Section 2.2.
The overall CDFS list has 396 X-ray sources out of which 302 have X-ray 
spectral fitting from \citet{tozzi06}.

For ECDFS sources outside of the CDFS, we use the X-ray catalog of 
\citet{lehmer05}, supplemented with currently available identification and
redshift information (Silverman et al. 2009, in preparation), but not yet 
including results of \citet{treister09}. Again, we 
have imposed a  $log(f_X/f_R)\geq -1$ cutoff
to reduce contamination by non-AGN. We avoid double counting objects
in the \citet{lehmer05} ECDFS catalog that are also CDFS sources.
Objects in the \citet{giacconi02} CDFS main catalog are identified 
by a flag
in \citet{lehmer05}. For these we use the CDFS data and spectral fitting only.
The \citet{luo08} data detect more sources at the edge of the CDFS 
that are in the \citet{lehmer05} catalog but not flagged as also detected in
the CDFS. We avoid 
double counting those by eliminating them via a 5\arcsec\ radius 
match and using the ECDFS X-ray data.

For the CDFS sources of \citet{tozzi06} we adopt the positions
provided in that paper. For the additional sources from \citet{luo08}
we use their optical positions where given and X-ray positions otherwise.
In the ECDFS, we use optical positions from identifications by Mainieri et al.
(in preparation) when available and X-ray positions elsewhere.
Note that the 27\arcsec\ FWHM beam of the LABOCA map, obtained after 
convolution of the raw map with its own beamsize of 19.2\arcsec\ 
\citep{weiss09}, is large 
compared to the \chandra\ X-ray positional uncertainities even for the least 
favourable case of modest S/N faint sources at large off-axis angles, 
reducing the need for optical identifications for the stacking 
process. Typically, \chandra\ X-ray positions are accurate to well 
below an arcsec \citep[e.g., Fig. 5 of][]{luo08}. 

We adopt as redshifts for all the CDFS sources of 
\citet{tozzi06} the values given in that paper. About 
half of these are spectroscopic 
redshifts from \citet{szokoly04} and other references, while the rest are 
photometric
redshifts which benefit from the excellent multiwavelength 
coverage of the CDFS. 
For the ECDFS area and the identifications of Mainieri et al., we have used 
in order of preference (1) secure (e.g., two or more emission lines or clear 
spectral features, Ca H\&K) spectroscopic redshifts from Silverman et al. 
(2009, in preparation), (2) other spectroscopic redshifts from the 
compilation of Rafferty et al. (2009, in preparation), (3) photometric 
redshifts derived by Rafferty et al. (2009, in preparation) from a 
comprehensive compilation of multi-band UV to \spitzer\/-IRAC photometry using 
the ZEBRA photo-z code and finally in a remaining 28 cases (4) photometric 
redshifts from the COMBO-17 survey which provides accurate photometric 
redshifts using photometry in 17 pass-bands from 350 to 930 nm.  Here we 
have used the 
latest version of the COMBO-17 CDFS catalogue, following a calibration update 
\citep{wolf08}. We use this dataset with two limitations. We
consider only COMBO-17 sources with R$<24$ (Vega): at these magnitudes
the errors on the photometric redshift estimates are expected to be
less than $|z_{\rm phot}-z_{\rm spec}|/(1+z_{\rm spec})\approx
0.06$. The COMBO-17 data for galaxies fainter than R$=24$ (Vega) are
too shallow for accurate photo-z determination of AGN. Further, we limit 
the use of COMBO-17 photometric
redshifts to z$<1.2$ because at higher redshifts the COMBO-17
estimates become increasingly inaccurate due to the lack of NIR
coverage (see Sec.  4.6 of \citet{wolf04}). We have waived this last 
constraint for objects
best fitted with a QSO templates ( MC$_{class} = {\rm 'QSO'}$ in
\citet{wolf04}) for which the photometric redshifts are accurate
at least to z$\approx 4$ (see Fig. 18 of \citet{wolf04}).
In total, we thus have redshifts for about three quarters of the X-ray targets 
outside
the CDFS. For the new \citet{luo08} sources we use redshifts from Rafferty 
et al. (2009, in preparation) where available.

The combined CDFS+ECDFS X-ray based catalog has 895 sources with a median 
X-ray flux of $3\times 10^{-16}\ergcms$ in the observed soft and  
$10^{-15}\ergcms$ in the hard band. 748 sources have spectroscopic or 
reliable photometric redshifts
in the range up to z$\sim$5 (median z=1.17), the median observed X-ray 
luminosity of those sources is $10^{43} \ergs$. In the following we call 
this the `combined' sample, while with `CDFS' sample we designate the 302  
sources from \citet{tozzi06} with X-ray spectral fits, remaining after the cuts
described above (median z=1.04). 

\subsection{Stacking procedure}

We derive average submm fluxes for a given source population by extracting both
the map flux (in Jy\,beam$^{-1}$) and the map rms noise, using bilinear 
interpolation
to the source position in the beam-convolved LABOCA flux map and noise map. 
We then use the inverse variance weighted average of the fluxes measured over
 the stack.

Stacking procedures for data from large-beam deep submillimeter maps that are 
approaching the confusion limit at these wavelengths have to consider two 
effects. First, blank background and thus the `zero point' of a sky image
is hard to identify. At the RMS noise of $\sim$1.2mJy of our data, integral 
number 
counts approach 10$^4$\,deg$^{-2}$ \citep[e.g.,][]{coppin06}, which is 
only a modest factor away from the LABOCA beam density ($2\times 
10^4$\,deg$^{-2}$ for an area of $\pi\,HWHM^2$ of the convolved beam);
see also the discussion in \citet{weiss09} on the contribution of confusion
noise in the LABOCA map.  This means  there is effectively no 
clean background sky. Second, for the large submm beam an elevated signal 
at the position of a stacked source may originate from the wings of the beam of
a nearby unrelated bright submm source. Chopped beam  patterns would cause 
additional complications but do not apply to the scanned LABOCA data that were
obtained in total power mode. 

\begin{figure}
\epsscale{1.}
\plotone{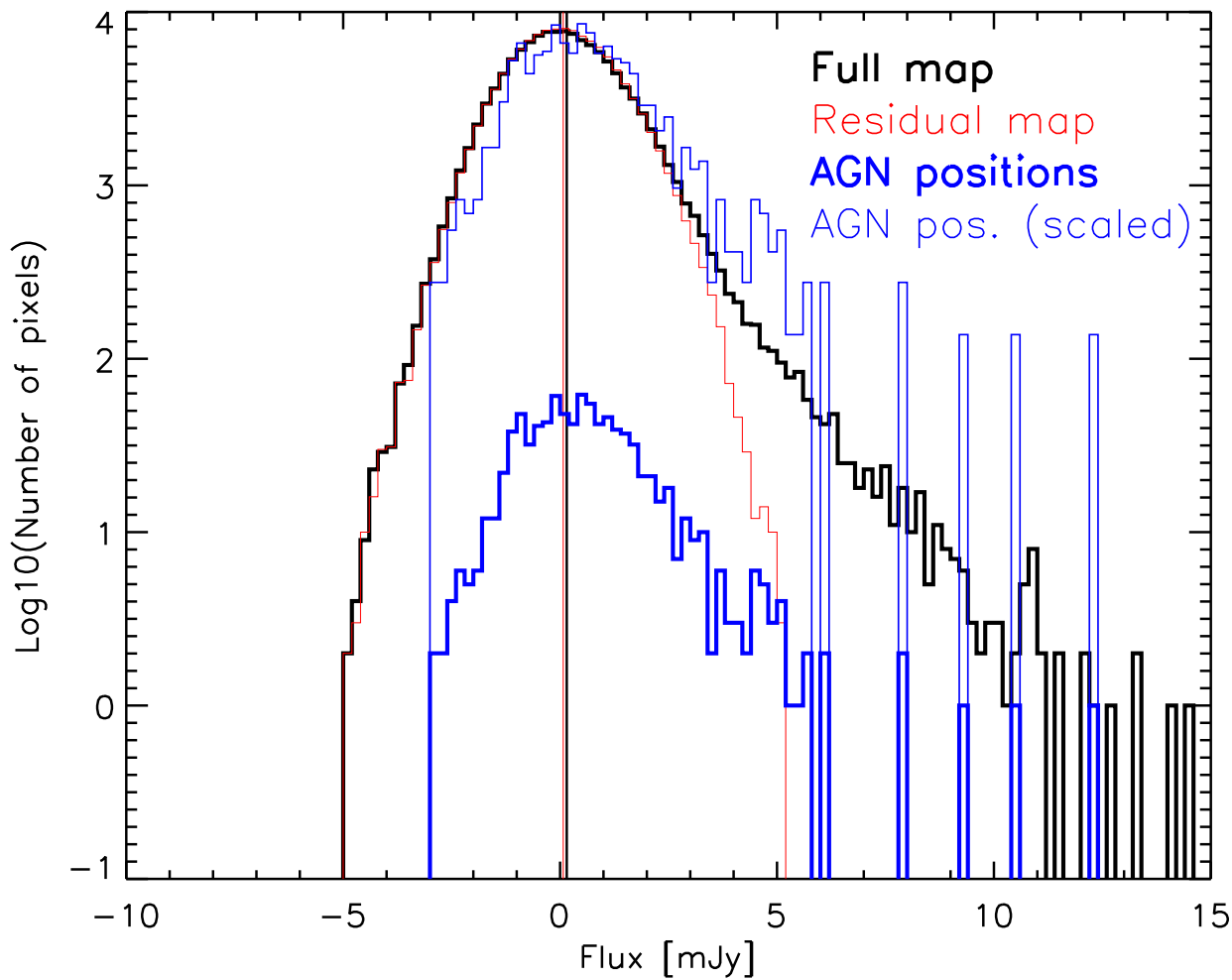}
\caption{Pixel flux histograms for the roughly 80\% of the LABOCA 
870$\mu$m map of the ECDFS with RMS 
less or equal 1.5 times the minimum noise of 1.07mJy. The black thick  
histogram represents the full map, the red thin histogram the residual 
map after subtraction of detected sources \citep{weiss09}. Vertical lines 
indicate the mean flux of this region in the respective map. The blue thick 
histogram shows the flux distribution at the positions of the combined 
sample of 895 CDFS and ECDFS AGN. The blue thin histogram shows for easy 
comparison this AGN histogram scaled to the number of map pixels.}
\label{fig:zero}
\end{figure}

We have addressed this situation with simple Monte Carlo simulations, 
randomly placing $\sim 10^6$ Gaussian beams over a 1000$\times$1000 pixel
blank image and assuming a beam width (in pixels) equal to the LABOCA beam 
width. Input fluxes were distributed according to a simplified integral 
number count 
distribution with a power law slope -2 \citep[approximating measurements of, 
e.g.,][]{coppin06,weiss09} and extending 3 orders of magnitude down
from the brightest source in the map, deep into confusion. If the simulated 
image is offset 
to a mean value 0 and stacking experiments are done using the positions of 
subsets of the input list, mean fluxes from stacks
agree with the mean of the input fluxes within an error estimated from the 
standard 
deviation of the image, divided by the square root of the stacked sample size.
The noise level of current submm maps and the fact that they do not reach 
below the knee of the integral counts implies a dynamic range 
between noise and brightest source of typically about an order of magnitude 
only. This reduces the effects of individual bright outliers on such error 
estimates. With a proper image zero point, the effects of the difficulty to 
define the background and of confusion thus cancel, and
stacked fluxes can be used directly. In the 
LABOCA data, instrument noise is still (just) dominant over confusion noise
\citep{weiss09}, its presence is again consistent with this stacking 
procedure. Fig.~\ref{fig:zero} shows the pixel flux 
histograms for the part of the LABOCA flux map that is within a factor 1.5
from the minimum RMS of 1.07\,mJy\,beam$^{-1}$ and for 
the residual map after subtraction of detected sources \citep{weiss09}.
In stacking experiments we subtracted the respective means 
(0.154\,mJy\,beam$^{-1}$ and
0.072\,mJy\,beam$^{-1}$ for this part of the flux and residual map, 
respectively). This
also ensures that zero average flux is returned from stacking random 
positions in the map. In deriving 
the stacked fluxes, we also calculate an inverse variance weighted stacked 
submap which, for stacks with significant detections, can
be used to verify that the stacked beam is centered and reproduces the 
original beam of the map.
Spatial clustering of the stacked population can in principle significantly 
affect source counts and stacking results. The effect on the stacks can be 
very important for instruments at similar wavelengths with much larger 
beams than that of APEX, like Planck-HFI 
\citep[e.g.,][]{negrello05}. It is negligible for the LABOCA beam 
at this wavelength \citep{bavouzet09}, and hence not considered
in our stacking.
  
We base our discussion on stacking the flux map but also provide stacking 
results 
for the residual map for comparison. In a few cases with poor statistics we 
explicitly  discuss the effects of significant detections for the stacked 
population. Given the noticeable overlap between the submillimeter population
and weak X-ray sources representing moderate luminosity AGN 
\citep[e.g.,][]{alexander03,alexander05b}, exclusion of individually
detected submm sources would bias our results. Future detailed identification 
campaigns and high spatial resolution submm followup of the LABOCA survey 
will allow supplementation of
this statistical approach with one based on the confirmed nature of individual 
sources.

We quote below the stacked fluxes for various samples of ECDFS AGN. In 
assigning errors we consider that, while the instrument noise 
is gaussian to good approximation \citep{weiss09}, the pixel histogram of
the LABOCA map is somewhat non-gaussian due to the effect of individually 
detected, as well as fainter, sources (Fig.~\ref{fig:zero}). Positive 
detections in stacks are hence more likely than for purely gaussian noise.
We provide in Tables~\ref{tab:stackcombo} and \ref{tab:stacktozzi} two error 
measures for the mean flux of each stack. For a sample with N sources, the 
error $\sigma_{map}$ provides the standard deviation from comparing the mean
fluxes of many ($>$1000) samples of N sources each, drawn at random spatial 
positions in the map. The value of  $\sigma_{map}$ allows assessment of the 
significance of a stack's detection. The error $\sigma_{subsample}$ provides 
the standard deviation 
from comparing the mean fluxes of many subsamples of N sources each, drawn
randomly from the fluxes measured at the positions of our combined 
sample of 895 AGN. 
This error is larger than $\sigma_{map}$ since it also includes the 
spread in the properties of the AGN population. For that reason, we use it 
when assessing the significance
of differences between subsamples of our overall AGN sample.
We have also compared $\sigma_{subsample}$ to error estimates from 
bootstrapping into each subsample proper and found the latter estimates 
broadly consistent, but with large fluctuations due to the sometimes small 
subsamples that are used below. 

\subsection{Stacking results for different AGN samples}

\begin{figure}
\epsscale{1.}
\plotone{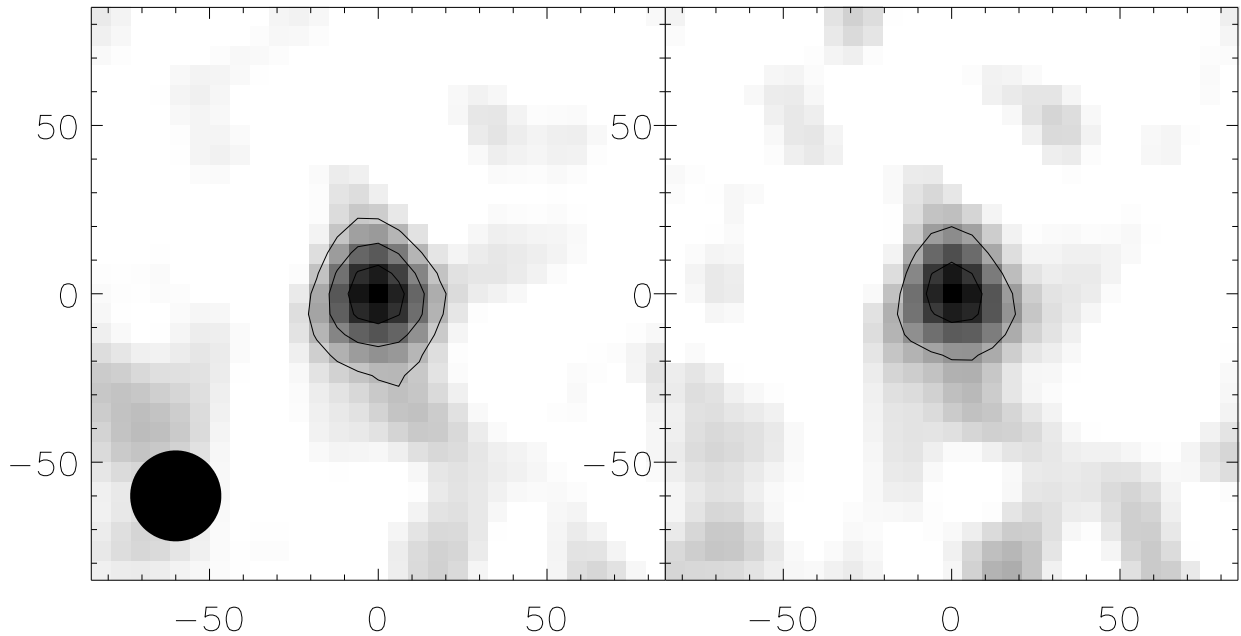}
\caption{Stamps showing the coadded LABOCA submm signal for all 895 CDFS and 
ECDFS X-ray sources in our combined list. Left: Obtained using the total 
LABOCA flux map. The 27\arcsec\ FWHM beamsize of the 
LABOCA map is shown for comparison in the bottom left. 
Right: Obtained using the residual map after subtraction of all 
sources individually detected above 3.7$\sigma$. The spatial scale is in 
arcseconds relative to the expected nominal center of the stacked beam. 
Contours are shown at 3, 6, 9 times the full stack noise of 0.044mJy. The 
greyscale of the images runs from zero to the respective peak.}
\label{fig:stamp}
\end{figure}

The stacked flux for the 895 X-ray sources from the combined CDFS+ECDFS
sample is detected clearly at S$_{870\mu m}$=0.49$\pm$0.044mJy (11.1$\sigma$).
The stacked image is shown in Fig.~\ref{fig:stamp}. A 
2-dimensional gaussian fit to the stacked beam is centered at 
$2.2\arcsec\pm 1.7\arcsec$
from the expected position
and has a gaussian fit FWHM of $27\arcsec\times 34\arcsec$ 
($\pm$3\arcsec). These parameters support the correctness
of the map reduction and the stacking procedure, given the nominal convolved
beam of 27\arcsec. Stacking the residual map after removal of all
$>3.7\sigma$ LABOCA point sources again provides a clear detection at 
$\sim$70\%\ of this 
flux (Fig.~\ref{fig:stamp}, Table~\ref{tab:stackcombo}). Use of the residual 
map will lower the average flux of a stacked X-ray population by excluding
members of that X-ray population that are individually detected SMGs. 
It can also lower the average flux if a subtracted point source is dominated 
by an unrelated object but includes a weaker submm flux that is originating 
from the blended X-ray source proper. Then, the flux at the position
of the nearby X-ray source will be oversubtracted by 
removing a point source with the combined flux. This is related to one part 
of the `boosting' effect on the fluxes of low S/N source detections from 
such a map. This is avoided for our residual map in a statistical sense 
because `deboosted' 
\citep{coppin05} fluxes for the detections have been used when deriving the 
residue \citep[see also][]{weiss09}.

For redshift z$\sim$1, close to the median redshift of (E)CDFS AGN, and 
adopting a T=35K, $\beta$=1.5 greybody far-infrared continuum 
shape for the star formation powered part of the SED, the total infrared 
luminosity corresponding to our 0.49\,mJy LABOCA
detection is $\sim 2.6\times 10^{11}\Lsun$. The inferred mean star 
formation rate is $\sim 27\Msun$\,yr$^{-1}$, assuming star formation 
dominated submm emission, the 
conversion of \citet{kennicutt98} and then
multiplying by 0.6 to convert to a \citet{chabrier03} initial mass function. 
Note that for z$>$0.5 and given the negative K correction for submm emission
\citep[e.g., Fig. 4 of ][]{blain02}, the considerable uncertainty
of this estimate is mostly in the adopted temperature of dust heated by 
star formation (Factor $\lesssim$2 in luminosity for a 5\,K difference), 
rather than in the difference between individual source redshifts and the 
redshift z=1 adopted for the conversion. The average deep X-ray 
field AGN is thus residing in a moderately actively star-forming object, 
its luminosity placing it in the category usually called luminous 
infrared galaxies (LIRGs). This result 
certainly averages over a range of far-infrared luminosities but the LABOCA 
detection also in the stacked residual map argues that it is not only due 
to a few luminous outliers. Better characterizing this spread will be a 
task for the \herschel\ Space Observatory, the $\sim$10mJy 
far-infrared SED peak expected for the adopted greybody is well 
within its capabilities. Given the $>11\sigma$ detection of the combined 
sample, stacks for subgroups can still be detectable at good significance. 
We use such substacks in the following to probe for trends with AGN 
properties.

\begin{figure}
\epsscale{0.6}
\plotone{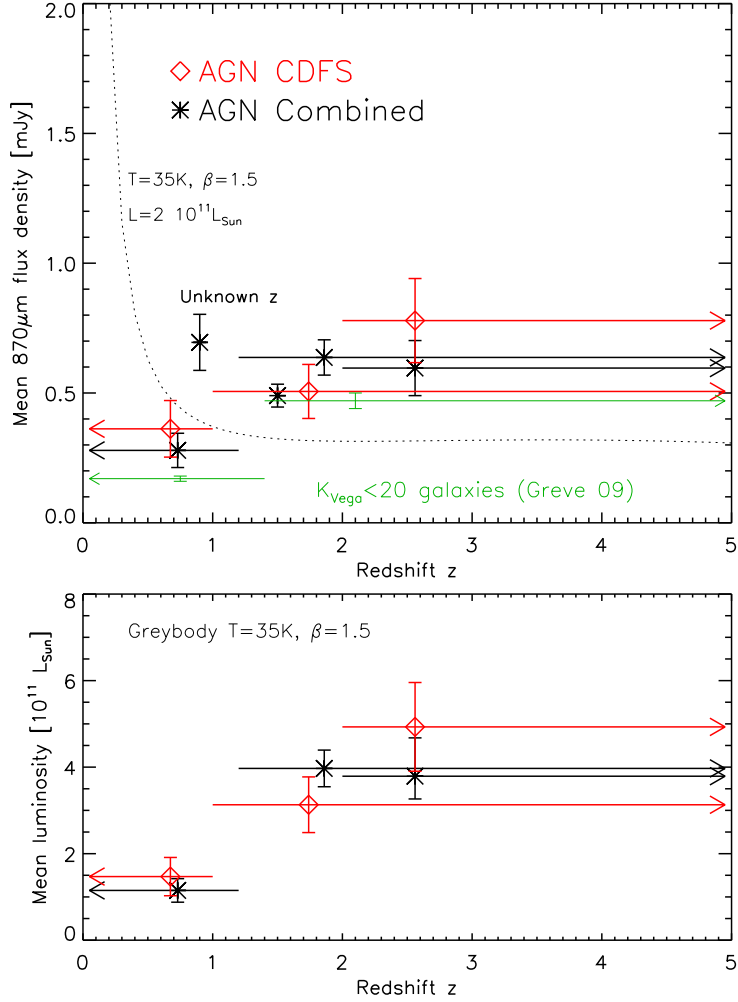}
\caption{Top panel: X-ray selected AGN from the CDFS and ECDFS show an 
increasing trend of mean submm flux density with redshift.  As in similar 
diagrams below, horizontal bars or arrows indicate the parameter range
covered by a subsample, while vertical error bars indicate the 1$\sigma$
errors of the mean flux for this subsample. The symbols are placed at the
median of the quantity on the abscissa for the respective bin - here redshift.
The asterisk without redshift range indicates the mean flux
of the entire combined sample. Results are shown separately for CDFS AGN 
only, and for the combined sample of CDFS and ECDFS AGN.
Results for ECDFS sources without reliable 
redshift are shown at an arbitrary z, most likely these contain a significant 
fraction of presently unidentified high redshift sources. The mean submm 
fluxes of (predominantly non-AGN) K$<$20 galaxies are added for comparison 
\citep[][also based on LABOCA data]{greve09}.
We also illustrate the submm flux expected from a greybody of given 
luminosity at different redshifts (dashed curve). Bottom panel:
Results from stacking the IR luminosities of greybodies fitted to the redshift
and map flux of each source.} 
\label{fig:ztrend}
\end{figure}

We start with a simple splitting of the combined sample by redshift 
(see Table~\ref{tab:stackcombo}
for results of this and subsequent splittings of the combined sample). More 
than 80\% of the 
sample have redshifts and are split into about equal groups below and 
above redshift z$=$1.2. There appears to be a trend towards higher 
star formation rates at 
higher redshift. A difference in mean submm flux between z$<$1.2 and z$>$1.2 
sources is found at the 3.0$\sigma$ level. As for other comparisons of AGN 
subsamples, we have adopted here the $\sigma_{subsample}$ errors  from 
Tables~\ref{tab:stackcombo} and \ref{tab:stacktozzi}. 
Comparing z$<$1.2 to z$>2$ sources gives a similar difference but only at 
the 2.1$\sigma$ level, due to the 
smaller size of the latter group. The sources without redshift assignment
on average have slightly higher submm flux than any of these groups, 
consistent with the notion that a significant fraction of them are 
located at high redshift and remain more difficult to identify 
\citep{alexander01,mainieri05a}. 
We can also scrutinize this trend with redshift via a Spearman rank 
correlation test of the 746 individual submm fluxes with redshift 
measurements, 
rather than looking at binned averages. The correlation coefficient is a 
modest 0.108, not surprising given the significant noise on each individual 
flux measurement, but the probability of exceeding this coefficient in the 
null hypothesis of uncorrelated data is only 0.003. Interpreting trends of
submm flux with redshift in terms of total infrared luminosity assumes their 
proportionality due to the negative K-correction. This assumption starts to
fail at z$\lesssim$0.5. However, changes in the ratio of submm flux and 
inferred infrared luminosity are small at the median redshifts of the 
bins discussed here 
(Fig.~\ref{fig:ztrend}). Deviations at z$<0.5$ act in the direction  of 
strengthening the trend of luminosity with redshift compared to the trend 
for submm flux. We illustrate this by reporting for those
samples in Table~\ref{tab:stackcombo} with complete redshifts also stacked
IR luminosities, obtained from the luminosities of fiducial T=35K, $\beta$=1.5
greybodies matched to redshift and submm flux of each source. These stacked
luminosities (see also lower panel of Fig.~\ref{fig:ztrend}) thus capture
the effect of the redshift distribution of the samples, and confirm its
impact to be small. Results from analysing redshift subgroups for
the CDFS sample only, with its more complete identification status 
(see Table~\ref{tab:stacktozzi} for this and other 
results for this sample), agree with results for the combined sample in 
showing higher 
flux at higher redshift but with $<2\sigma$ significance of the trends 
at the given sample sizes. Of course, the trends 
with redshift (see also Fig.~\ref{fig:ztrend}) do not immediately imply 
evolution, because of the luminosity vs. redshift selection effects of 
the underlying X-ray surveys which are effectively flux limited.

\begin{figure}
\epsscale{0.6}
\plotone{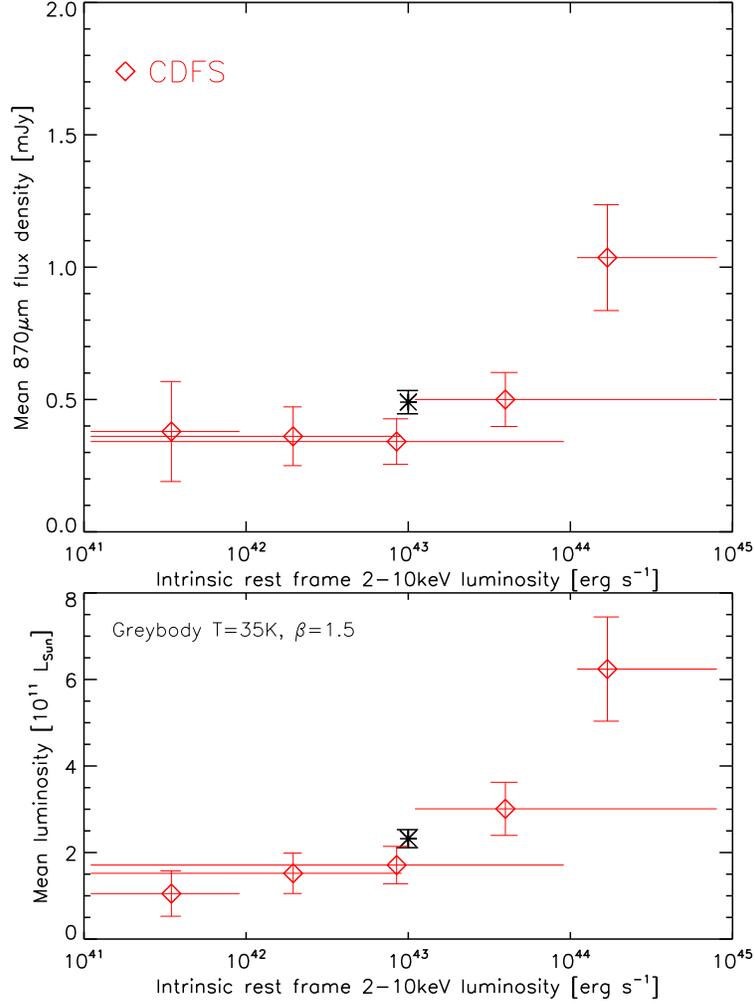}
\caption{Top panel: Mean submm flux density as a function of  intrinsic 
rest frame 2-10keV
luminosity of the AGN. The asterisk indicates the mean submm flux of the 
entire combined sample. The mean submm flux rises clearly above 
$L_{2-10keV}=10^{44}\ergs$. These highest luminosities are also the sole driver
of the flux increase in the wider $L_{2-10keV}=10^{43}\ergs$ bin. 
Bottom panel: Results from stacking the IR luminosities of greybodies 
fitted to the redshift and map flux of each source.}
\label{fig:lumtrend}
\end{figure}

To test this further, trends with X-ray luminosity need to be explored.
Roughly half of those sources in the combined sample with redshifts are not 
used here, because of the lack of X-ray spectral modelling for many of the 
ECDFS-only sources. For them, only a simple
observed X-ray luminosity can be computed from the distance and observed 
X-ray fluxes. No significant trends of average submm flux can be observed
with this observed X-ray luminosity, but this could also be due to the effect 
of variations in the obscuring column density erasing trends with 
intrinsic AGN luminosity.
This aspect is better addressed using only the CDFS sample, for which the 
spectral fits of \citet{tozzi06} assign an intrinsic rest frame
hard X-ray (2-10\,keV)
luminosity for all sources including hard/obscured ones. We find no 
noticeable differences between groups or with respect to the total sample 
when dividing at $L_{2-10keV}=10^{43}\ergs$ into 
a more and a less luminous group with about half of the sample each  
(Table~\ref{tab:stacktozzi}, Fig.~\ref{fig:lumtrend}). This changes at 
the most luminous
end: Sources with $L_{2-10keV}>10^{44}\ergs$ have more than twice
the average submm flux, and differ from the average of the CDFS sample 
at 2.2$\sigma$ and
from the sources with $L_{2-10keV}<10^{44}\ergs$ at 2.6$\sigma$.
We further discuss this stronger star formation around the most luminous
X-ray AGN below. Again the behaviour implied by the analysis of
the stacks can be tested via a rank 
correlation of individual submm fluxes with log($L_{2-10keV}$). Given 
the upturn in submm only at the high X-ray luminosity end 
(Fig.~\ref{fig:lumtrend}) 
it is not suprising that no significant correlation is seen over the 
full range (N=302, C.C.=0.059, Significance 0.30), but correlation between
submm flux and luminosity is found above $L_{2-10keV}>10^{43.5}\ergs$
(N=98, C.C.=0.26, Significance 0.01).

\begin{figure}
\epsscale{.7}
\plotone{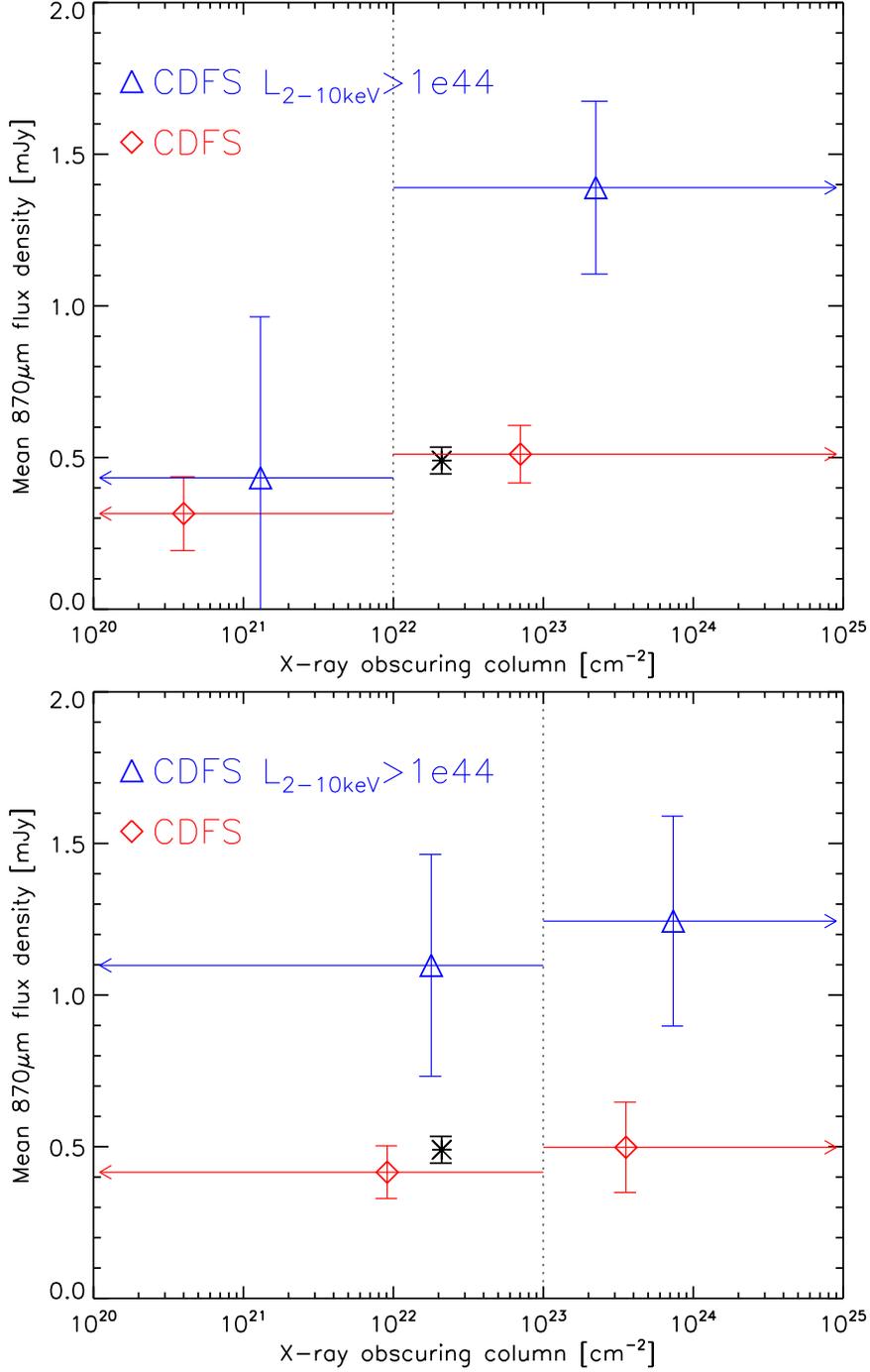}
\caption{Trends of mean submm flux density with AGN X-ray obscuring column
density. 
Top panel: Sample split at $N_{H}=10^{22}\cmsq$, as indicated by the dotted
line. Bottom panel: Sample split at $N_{H}=10^{23}\cmsq$. 
 The asterisk shows the mean flux of the entire combined sample. We
show results for the full CDFS sample, with no indication for a trend with 
$N_{H}$, and for the the most X-ray luminous CDFS sources with good
redshifts. From the top panel, these may be more submm luminous 
above $N_{H}=10^{22}\cmsq$ compared to below, but the 
difference is not significant in our sample.}
\label{fig:nhtrend}
\end{figure}

One of the most interesting parameters that is potentially linked by evolution 
to the level of star
formation is the X-ray obscuring column density. Simple tests that 
can be done on the combined sample show no trend: Binning
directly by X-ray hardness ratio as well as by crude estimates of the 
obscuring column density from the location in the redshift vs. hardness ratio
diagram \citep[e.g., Fig.~8 of][]{szokoly04} does not show any noteworthy 
changes. We hence focus again on the CDFS sample using the modelled 
X-ray obscuring column densities $N_H$ from \citet{tozzi06}. As 
Table~\ref{tab:stacktozzi} and Fig.~\ref{fig:nhtrend} show, there is no 
significant variation with obscuring column density even for this 
well-characterized sample and comparing subsamples that are individually 
detected at 3--5$\sigma$. We here split the sample at two different column
densities. First, $N_{H}=10^{22}\cmsq$ is often used as the limit 
distinguishing unobscured AGN from the larger number of obscured AGN 
\citep[e.g.,][]{tozzi06}. As a second test we 
broke the sample at $N_{H}=10^{23}\cmsq$ which would help identifying 
changes for the highest column density objects. These two cuts, as well as an 
intermediate one at  $N_{H}=3\times 10^{22}\cmsq$ that gives roughly 
equally populated bins above and below the threshold, do not reveal any 
significant trends with X-ray obscuring column density. For the 
$N_{H}=3\times 10^{22}\cmsq$ cut which has the most equally distributed 
statistics, the ratio of stacked submm fluxes for sources more/less 
obscured than the cut is $1.1\pm 0.5$, well below high ratios like
the factor 4.4 found in the \citet{stevens05} comparison of obscured
and unobscured very luminous QSOs. Using submm emission as star formation
indicator, there hence seems to be no clear trend of host star formation with
nuclear obscuration for the typical $L_{2-10keV}\sim 10^{43}\ergs$ AGN.
Recent attempts using radio continuum emission \citep{rovilos07} and the 
[O{\sc II}]$\lambda$3727 emission line \citep{silverman09} to trace star 
formation agree in not finding such trends.

\subsection{Results for the most luminous CDFS X-ray AGN}

In the previous section, we have found for the full population of (E)CDFS X-ray
AGN a trend of submm flux and star formation rate with redshift, a change 
with intrinsic X-ray luminosity only at the highest 
$L_{2-10keV}\geq 10^{44}\ergs$
AGN luminosities, the border that is conventionally adopted between 
`AGN' and `QSO', and no significant change with X-ray 
obscuring column density. We will discuss the implications below but first
specifically repeat the check for possible trends with obscuration at the
highest AGN luminosities. This is motivated by the large range of luminosities
covered, combined with the possibility that evolutionary paths may 
significantly differ between luminous QSOs and lower luminosity AGN.  

We first restrict the stacks which are formed via the \citet{tozzi06} 
obscuring column densities to AGN with high intrinsic luminosity 
$L_{2-10keV}\geq 3\times 10^{43}\ergs$. A column density
$N_{H}=10^{23}\cmsq$ divides the sample roughly into half, but we also
explore the traditional $N_{H}=10^{22}\cmsq$ separation. 
In both cases the more obscured sources appear brighter in the submm
(Table~\ref{tab:stacktozzi}),
but the difference stays well below 2$\sigma$ significance in either case.
As a next step toward the most luminous and reliable sources we
further restrict to $L_{2-10keV}\geq 10^{44}\ergs$
and sources with spectroscopic redshifts (redshift quality $\geq$1 in 
\citet{tozzi06}). Again we find the more obscured sources to be submm 
brighter (Table~\ref{tab:stacktozzi}, Fig.~\ref{fig:nhtrend}) with again a 
larger difference in submm flux if separating at $N_{H}=10^{22}\cmsq$, but low 
significance. Such a change in mean star forming rate with 
obscuring column density
occuring at roughly $N_{H}=10^{22}\cmsq$ might indicate that the
small variations with column density for the full CDFS sample 
may be dominated by the contribution of luminous AGN. Consistent with this
interpretation, a test splitting the
less luminous $L_{2-10keV}< 3\times 10^{43}\ergs$ AGN at 
$N_{H}=10^{22}\cmsq$ further reduces the difference in submm flux 
between the two bins, compared to the full sample 
(Table~\ref{tab:stacktozzi}).

An independent way of distinguishing unobscured from obscured AGN is the 
optical spectral classification. We use the highest luminosity AGN from
\citet{tozzi06}, for which either $L_{0.5-2keV}$ or $L_{2-10keV}$ is at least 
$10^{44}\ergs$. Among those, we restrict ourselves to sources with
spectroscopic redshift (redshift quality flag $\geq$1 in \citet{tozzi06}) 
and group them into optical Type 1 (BLAGN in the scheme of 
\citet{szokoly04}, 10 sources\footnote{\citet{giacconi02} XIDs 11, 22, 24, 
42, 60, 62, 67, 68, 91, 206})
or Type 2 (HEX, LEX or ABS classifications in the scheme of 
\citet{szokoly04}, 17 sources\footnote{\citet{giacconi02} XIDs 18, 27, 
31, 35, 45, 51, 57, 76, 112, 153, 
156, 202, 253 (changed redshift from \citet{szokoly04} in \citet{tozzi06}, 
see also \citet{roche06}), 263, 268, 547, 601}). Here we included XID 35 
that was not classified by \citet{szokoly04} but for which an optical 
spectrum from the VVDS survey \citep{lefevre04} is lacking obvious 
broad lines.
Consistent with the analysis where obscuration was defined via the
X-ray obscuring column density, the obscured optical Type 2 AGN 
are again submm brighter (Table~\ref{tab:stacktozzi}) than the unobscured 
Type 1 AGN, but the significance of the difference is low at just 1$\sigma$.

For both methods, the difference between submm fluxes of 
luminous unobscured and obscured AGN is stronger when stacking the flux 
map compared to stacking the residual map, suggesting a contribution of 
individually detected submm galaxies.
Inspection shows that the `hard' group defined by $N_{H}>10^{22}\cmsq$ or
by optical Type 2 in both cases includes two X-ray sources with S/N in the
submm map above 3 at their position, because they are close to
submm sources that are detected at $>3.7\sigma$. Both have extracted submm 
fluxes of $\gtrsim$6mJy. The `soft' group in contrast has no such sources. 
These two
SMGs are near the X-ray sources with \citet{giacconi02} XIDs 51 (z=1.097) 
and 112 (z=2.940). We have investigated these associations on the LABOCA 
map, the VLA data of \citet{miller08} and the FIDEL 24$\mu$m image 
(M. Dickinson et al., in preparation). For XID 112,
the position of the optical identification for which \citet{szokoly04} 
determined the redshift agrees within 1\arcsec\ with a weak 
S$_{1.4GHz}$=48$\mu$Jy radio source and a 24$\mu$m source, and 
is only 2.4\arcsec\ from SMG
LESS J33152.0-275329 \citep{weiss09}. This supports the association 
despite the fact this is a relatively complex region of the LABOCA map 
with two more submm sources within less than 1 arcmin.
For XID 52, again the optical position agrees with a (93$\mu$Jy) radio source 
and a 24$\mu$m source, but the offset to LESS J33217.6-275230 is 10.8\arcsec,
large ($\gtrsim 2\sigma$) for identifications of such a $\sim$5$\sigma$ submm 
detection and the 27\arcsec\ convolved LABOCA beam 
(e.g., equation B22 of \citet{ivison07} and the 6\arcsec\ positional accuracy 
estimated for the LABOCA map in \citet{weiss09}). We maintain this source in 
our stacks given that there is no nearby alternative radio identification 
for the SMG, 
and the region is again complex with the next SMG detection almost blended, 
but this association is clearly uncertain. 

In our comparison of obscured and unobscured 
$L_{2-10keV}>10^{44}\ergs$ AGN we have found brighter submm 
fluxes by a factor $\sim$3 comparing X-ray column densities above to below 
$N_{H}=10^{22}\cmsq$ and by a factor $\sim$2 comparing optical Type 2 
to Type 1. This is intriguing but the significance of the differences is 
too low in either case to claim a detection from our sample. 
 The next steps in sample and field size and/or errors on the 
individual star forming rates will be needed for robustly confirming 
whether the stronger star formation reported 
at higher AGN luminosity in X-ray obscured vs. unobscured broad-line AGN
\citep[][and subsequent work]{page01} also holds in the regime
of the brightest X-ray sources found in deep surveys like the (E)CDFS.

\subsection{Results on individual literature X-ray sources}
CXOCDFS J033229.9-275106 \citep[also called CDFS-202;][]{norman02} at 
redshift of z=3.70 has been considered a prototype of a luminous radio-quiet 
X-ray selected type 2 QSO. \citet{sturm06} have obtained a deep \spitzer\ 
mid-infrared spectrum of this source, detecting AGN continuum but not the
6.2$\mu$m rest wavelength PAH emission feature that could be detectable at 
this depth 
if the source were also hosting an extremely luminous SMG-like starburst. In
accordance with this result, the LABOCA map does not show a detection
at the position of CDF-S 202 (S$_{870}=-1.01\pm 1.10$mJy).
  
From deep SCUBA follow up of four heavily
X-ray absorbed and X-ray luminous AGN in the CDFS, \citet{mainieri05} report a 
S$_{850}=4.8\pm 1.1$mJy detection for the z=3.66 object CDFS-263.
The LABOCA map gives a flux of S$_{870}=1.88\pm 1.16$mJy for this source.
While we cannot support the \citet{mainieri05} result by an independent
significant detection, the two measurements are still consistent within 
2$\sigma$ given the errors, and the 
source is likely among the submm brighter part of the X-ray AGN population.
\citet{rigopoulou09} publish results from an extension of this project to 8 
sources. They do not detect at 850$\mu$m CXOCDFS J033229.9-275106, consistent 
with the LABOCA and Spitzer results reported for this source above. 
Stacking their 8 targets in the LABOCA map we find 
S$_{870}=1.55\pm 0.41$mJy, consistent with results for 
$L_{2-10keV}>10^{44}\ergs$ Type 2 AGN reported above but somewhat lower than 
the S$_{870}=4.0\pm 0.5$mJy obtained from scaling the variance-weighted 
mean of the \citet{rigopoulou09} SCUBA results to 870$\mu$m.

\citet{koekemoer04} report the detection in the GOODS-S region of 7 
`extreme X-ray/optical ratio sources' (EXOs) characterized by robust 
\chandra\ X-ray detections but optical nondetections to extremely low limits.
All of these fall on the LABOCA map, none of them is individually detected and
the stacked flux is 0.07$\pm$0.43mJy. While this nondetection at submm 
wavelengths excludes an explanation of EXOs by obscured AGN that are 
coexistent with extreme star formation, it is 
compatible with other possible spectral energy distributions of AGN at 
moderate to extremely high redshifts.

\section{Discussion}

Our data put strong limits on possible trends in submm brightness with 
obscuration for AGN with moderate luminosities, but are consistent with such 
a trend for luminous AGN. This result can be compared to previous
studies. The first comparisons of deep X-ray and submm surveys mentioned
in the introduction had too limited statistics to reliably address this issue.
\citet{page01} used a different approach of selecting very luminous 
hard X-ray sources from the \rosat\ survey and
obtained significant submm detections in 4 out of 8 z=1--2.8 obscured AGN,
the weighted mean submm flux for all eight is S$_{850}=4.4\pm 0.5$mJy. 
Subsequent papers
\citep{page04,stevens05} supported this result by matching to 
unobscured AGN of similar luminosity and redshift, and by extension of sample 
size, overall finding a ratio 4.4 in submm flux
between X-ray unobscured and obscured AGN,
and with the difference between submm detection rates of individual
objects in the two groups significant at the 3-4$\sigma$ level.  

Our LABOCA results grouping luminous ($L_{2-10keV}>10^{44}\ergs$) 
CDFS X-ray AGN 
by obscuration provide a ratio 2--3 in submm flux between obscured and 
unobscured objects, using separations by  either $N_{H}=10^{22}\cmsq$ or
by optical spectral type. However, due to the low significance we can not rule
out neither absence of a difference nor trends as reported by 
\citet{page01} and subsequent papers. One should note that, considering 
the $\sim$8\%\ correction from 870$\mu$m to 850$\mu$m flux densities 
for $\beta$=1.5 optically thin dust at z=1--4, our average 
submm fluxes of obscured sources are a factor $\sim$2-3 lower than reported by
\citet{page01} and \citet{stevens05} for their samples. Assuming that the 
increase of submm
flux with X-ray luminosity at highest luminosities (Fig.~\ref{fig:lumtrend})
continues beyond the range  up to $L_{2-10keV}\sim 8\times 10^{44}\ergs$ 
covered by our 
sample, this difference may be due to the typically higher  
$L_{2-10KeV}\sim 10^{45}\ergs$ luminosities of the \citet{stevens05} sample
(see also Fig.~\ref{fig:growthplot}). Observations with better statistics
at $L_{2-10keV}\gtrsim 10^{44}\ergs$ will be needed to firmly establish 
whether in this regime not only the observed increase in submm flux occurs, 
but also a gap starts to open between submm properties of unobscured and 
obscured AGN.

An important point to note is that the \citet{stevens05} obscured QSOs are 
X-ray absorbed (though with typical  $N_{H}$ just above $10^{22}\cmsq$ not 
Compton thick) but optical broad line region (Type 1) objects. Stevens et al. 
are comparing optical Type 1 QSOs having low X-ray absorbing column density
with Type 1 but higher X-ray absorbing column density. In contrast, we were 
first comparing groups defined by X-ray column density only, and in the 
second test comparing optical 
Type 1 to optical Type 2. Splitting our very small sample of 10 
X-ray luminous BLAGN further by X-ray column density to fully reproduce the
\citet{stevens05} approach did not show significant differences of subgroups in
this group. While our result is consistent with
the \citet{stevens05} finding of
increased submm flux for obscured luminous AGN, the detailed role of different
definitions of AGN obscuration remains to be investigated with better 
sensitivity and statistics. 

\citet{sturm06} used \spitzer\ mid-infrared spectroscopy to detect or put
limits on star formation in a sample of eight optical type 2 QSOs selected 
from deep X-ray surveys. These objects span a range of
intrinsic luminosities $L_{0.5-10keV}=10^{43.1}$ to $10^{45}\ergs$ and 
X-ray obscuring column densities  
$N_{H}=10^{21.3}$ to $>10^{24}\cmsq$, with all but one above 10$^{22}\cmsq$. 
We discussed the consistent LABOCA results for the z$\sim$3.7 source CDFS-202 
above. For the major part of their sample, six sources at redshifts
0.205 to 1.38 (median $\sim$0.5), \citet{sturm06} report one detection and five
limits on mid-infrared PAH emission with inferred star forming 
luminosities of the order $2\times 10^{10}$\Lsun. This is
lower than the $\sim1.1\times 10^{11}$\Lsun\ obtained from the 
mean LABOCA flux for the z$<$1.2 redshift bin, averaging over  all X-ray 
luminosities (Fig.~\ref{fig:ztrend}), converting via 
our adoted T=35K, $\beta$=1.5 SED and assuming z=0.6. The \citet{sturm06} 
sample is small and may have missed star forming objects, but another factor 
likely contributing to the difference is the extrapolation from the 
PAH measurements to total star forming luminosity which is a function of
interstellar medium conditions. \citet{sturm06} adopted a 
scaling factor based on the star forming galaxy M82. A scaling factor more 
similar to the one for high radiation field intensity environments, as 
suggested for star formation in hosts of  local type 1 QSOs by 
\citet{schweitzer06}, would increase the inferred star formation 
rates/limits of \citet{sturm06} by a factor $\sim$4 and 
bring them close to the typical z$<$1 star formation rate inferred from the
LABOCA submm fluxes.

Another modestly sized z$\sim$0.5 QSO2 sample studied spectroscopically 
with \spitzer\ 
was presented by \citet{zakamska08}. They observed 12 type 2 QSOs, 10 of which
were selected from a large area SDSS sample primarily based on their optical 
Type 2 
spectra, large [OIII] luminosities and bright mid-infrared continua. Two 
sources of their sample of 12 entered the sample 
from different far-infrared selected programs. Compared to the Sturm et al. 
QSO2s they exhibit larger AGN mid-IR luminosities and, on the basis of X-ray
to [OIII] ratios, possibly higher X-ray obscuring column densities. 
Zakamska et al. report PAH detections in 6 of 12 objects and infer a typical
star formation luminosity of  $\sim 5\times 10^{11}$\Lsun from the median of
detections and limits. Both of these \spitzer\ spectroscopic studies are 
broadly consistent with the LIRG-like star forming luminosities of luminous
type 2 AGN found in our LABOCA study. The overlap in luminosity, redshift 
and other properties is not large enough for a direct comparison of methods. 
A combination into a single analysis is not straightforward, given the 
different uncertainties that are involved in extrapolation to far-infrared 
luminosity from either the mid-infrared or submm side.

\subsection{The co-evolution of X-ray survey AGN with their hosts: 
Two paths?}

The differences expected in the merger evolution scenario between star 
formation around obscured and unobscured AGN may be present only at 
the high luminosity end of our sample and have been reported by 
\citet{page01} at yet higher luminosities. For the 
bulk of lower luminosity $L_{2-10keV}\sim 10^{43}\ergs$ sources 
we do not observe any significant trend with obscuration. This is likely 
indicating a different evolutionary path for these AGN and their hosts.

The host properties of high-redshift Type 1 and luminous Type 2 AGN are 
difficult to constrain via common optical/near-infrared techniques, as
the AGN often outshines the host. For the more accessible intermediate 
luminosity and Type 2 part of the population, several studies have concluded
that z$\lesssim$1 X-ray AGN are hosted by massive galaxies spanning the 
region from around the top of the `blue cloud' via the `green valley' to 
the `red sequence' in a color - absolute magnitude diagram 
\citep{nandra07,silverman08,treister09}. These are luminous host galaxies, 
with few of them fainter than rest frame absolute magnitude 
$M_B=-20.5$ and typically brighter than $M_B=-21$ \citep[see also][]{barger03}.
Photometric stellar mass analyses for z$\sim$1 X-ray AGN hosts 
\citep{alonso08,bundy08,lehmer08,silverman09} and analogy to the 
$M_B$ -- stellar mass relation for the general z=0.7-1 population 
\citep[e.g.,][]{cooper08} suggests log(M*)$\gtrsim$10.5\Msun\ massive hosts. 
It is  plausible from these results as well as from the local evidence 
\citep{kauffmann03} to assume that also the bulk of the z$\sim$1 
unobscured/luminous X-ray AGN population resides in massive host galaxies. 
 
Several studies have recently used mainly mid-infrared star formation 
indicators to conclude that typical massive galaxies at redshifts 0.7--2 are 
almost constantly forming stars at considerable star formation rates,
with the stellar mass normalized `specific star formation rate' (SSFR) 
increasing with redshift. This is concluded from the presence of a fairly 
tight mass -- SFR relation which shifts towards higher SSFR with redshift
\citep{noeske07,elbaz07,daddi07}. The tightness of this relation suggests a 
high duty cycle of star formation. For massive z$\sim$ 2 galaxies, a strong 
role of secular evolutionary 
processes compared to individual brief merger events is independently 
suggested by dynamical studies of rest frame optical/UV selected high 
redshift galaxies  
\citep[e.g.,][]{foerster06,genzel08,shapiro08}.

This developing picture of z$\sim$1--2 galaxy evolution is naturally
complemented by our results for the bulk of the X-ray sample: Strong 
trends of host star formation rate with AGN obscuration are lacking 
because such star formation rates are pervasive in galaxies of 
similar mass 
and redshift. The hosts are evolving secularly and star formation is not
linked to a specific state of the AGN. The typical $\sim$30\Msun\,yr$^{-1}$ 
estimated above for AGN hosts assuming star formation dominated submm emission
compares well with typical star formation rates in z$\sim$1
log(M*)$\sim$10.5\Msun\ galaxies \citep{noeske07,daddi07}. A further 
comparative interpretation is currently not warranted given the 
uncertainty in the stellar  masses 
of the AGN hosts and the fact that the comparison of star formation rates 
is subject to different extrapolation effects.
Star formation rates in these studies  are based on extrapolation from the 
optical/mid-infrared while we extrapolate from the submm. This mismatch 
can be partly remedied by a comparison to submm fluxes (measuring the rest 
frame far-infrared and indirectly star 
formation rates) that were estimated for optically
selected galaxies from the LESS LABOCA survey. \citet{greve09} find for a 
K-selected K$_{Vega}\leq 20$ sample stacked 870$\mu$m fluxes of 
0.17$\pm$0.01\,mJy 
(z$<$1.4) and 0.47$\pm$0.03\,mJy (z$>$1.4), similar but slightly below our 
trend for the AGN hosts, and again with a positive trend with redshift. 
The increase in AGN host star formation rate with redshift 
(Fig.~\ref{fig:ztrend}) is combined with a relation SFR to AGN luminosity 
that is flat over a wide range of moderate AGN luminosities 
(Fig.~\ref{fig:lumtrend}), i.e. SFR does not depend on AGN luminosity. 
This is consistent with the moderate luminosity AGN hosts
indeed following the increase of SSFR with redshift of the general galaxy
population. Firmly establishing this behaviour will require studies with better
SFR sensitivity for individual objects, removing the need to average over 
large stacks and better breaking redshift-luminosity correlations in the parent
X-ray sample. \citet{silverman09} use the AGN-subtracted 
[O{\sc II}]$\lambda$3727 emission line to compare star formation rates 
in $0.48<z<1.02$ AGN and inactive galaxies of same stellar mass at same 
redshift. For moderate luminosity  $42<log(L_{0.5-10keV})<43.7$ they find 
indistinguishable SFR distributions in full agreement with the submm result.
When including larger AGN luminosities a SFR excess is indicated in their 
data. If robust to the increasing technical difficulties of measuring mass
and [O{\sc II}] SFR for luminous AGN, this could indicate the onset of star
formation enhancement due to e.g. merging.

Our interpretation that most of the moderate luminosity X-ray survey AGN are 
hosted by massive secularly evolving galaxies is consistent with morphological
analyses. \hst\ studies of z$\sim$1 X-ray AGN hosts \citep{grogin05, pierce07}
find the hosts to typically be bulge-dominated galaxies with only a modest 
fraction of hosts showing clear morphological evidence for recent major 
mergers, such as strong asymmetries.

Analysis of deep X-ray surveys has clearly shown a difference in the redshift 
evolution of high and lower luminosity AGN, with the comoving  density of 
lower luminosity AGN peaking at lower redshift than the z$\sim$2 `Quasar 
epoch'. Since cosmic halo merger evolution can be roughly matched to 
the evolution of quasars but not to the evolution of lower luminosity AGN, 
this has been interpreted in terms of a difference between
merger driven accretion at high AGN luminosities and more secular evolution
at lower luminosities \citep[e.g.,][and references therein]
{hasinger08, hopkins09}.

All these lines of evidence place the bulk of the AGN population detected
in deep X-ray surveys on
a relatively gentle and secular evolutionary path. In these sources, X-ray
obscuration may vary through orientation of the immediate AGN environment
in a classical unified/torus picture, and possibly also with an additional 
contribution by obscuration on larger host scales, but obscuration is not 
intimately linked to the 
global star formation rate of the host. Only for the  
$L_{2-10keV}\gtrsim 10^{44}\ergs$ QSO-like AGN, obscuration may coincide 
with high star formation, consistent with a classical merger evolutionary path.

\subsection{Star formation and AGN accretion rates}

Over cosmic time, the growth and merger rates of black holes and their hosts 
have to establish the local black hole -- bulge mass relation 
\citep[e.g.,][]{marconi03,haering04}. In this context, it 
is interesting to compare for high redshift AGN populations the current
accretion rate onto the black hole and the host growth via star formation
 with a relation that would establish on average the local ratio of black hole
and bulge mass. Our results permit steps in this direction for X-ray selected
AGN, within the constraints of results that are based on 
averaging over sizeable samples.   

\begin{figure}
\epsscale{.7}
\plotone{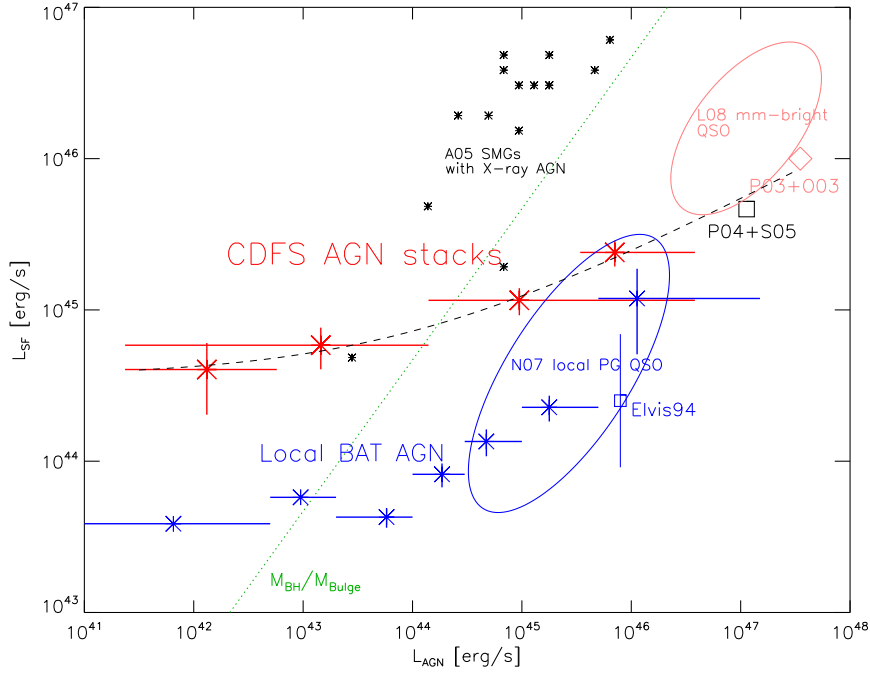}
\caption{AGN luminosities and star forming far-infrared luminosities for 
high redshift and local AGN. Moderate luminosity AGN have host star formation
rather independent of AGN luminosity, but at a level rising with redshift.
For high AGN luminosity, AGN luminosity and host star formation correlate. 
Red asterisks represent the four CDFS AGN stacks grouped by 
intrinsic 2-10keV X-ray luminosity, as also shown in Fig.\ref{fig:lumtrend} 
and Table~\ref{tab:stacktozzi}. 
The four stacks are for $L_{2-10keV}<10^{42}$, $<10^{43}$, $>10^{43}$ and
$>10^{44}\ergs$ (see also the horizontal bars indicating the range of AGN 
luminosities entering) and are plotted at the median AGN 
luminosity of the sources in each stack.
The diagonal dotted green line indicates the relation for a 
continuous host and black hole growth, reflecting the local universe relation 
between black hole mass and bulge mass. The black square represents the 
combined X-ray obscured and unobscured samples of more luminous X-ray AGN from
\citet{page04} and \citet{stevens05} [P04, S05], see also Section 3. AGN 
and star formation luminosities have been obtained analogously to the CDFS
stacks. The pink ellipse indicates the location in such a diagram of 
mm-bright and optically
very luminous high-z QSOs studied by \citet{lutz08} [L08]. The pink diamond 
reflects the suggested median location of the parent population of these 
luminous high-z QSOs including mm-faint ones, as discussed in \citet{lutz08} 
and based on the surveys of \citet{priddey03} and \citet{omont03} [P03, O03]. 
The dashed
line is a simple constant plus power law relation between AGN and star forming
luminosity approximating the high redshift AGN. We also include local 
universe (z$<0.3$) AGN, the blue ellipse representing local PG QSOs studied 
by \citet{netzer07} and the blue asterisks stacks of local hard X-ray
selected AGN based on \swift\/-BAT and \iras\ data (Section 3.3). The blue 
square represents the \cite{elvis94} QSO sample at its median AGN luminosity
converted to our cosmology, with the error bar indicating the 1$\sigma$ 
variation in FIR/optical (star formation/AGN) flux ratio of objects in that 
sample. Submillimeter galaxies with weak X-ray AGN \citep{alexander05a} are 
shown for comparison.
}
\label{fig:growthplot}
\end{figure}

In Fig.~\ref{fig:growthplot} we show the location of the CDFS stacks, grouped 
by intrinsic 2--10keV X-ray luminosity as in Fig.~\ref{fig:lumtrend}, now in 
a diagram comparing star forming and AGN
luminosity. To obtain the AGN luminosities we have converted from median 
2-10keV X-ray luminosity for each luminosity bin, assumed to be AGN dominated,
to monochromatic 5100\AA\ luminosity using the luminosity-dependent 
$\alpha_{OX}$ relation of \citet{steffen06} as cast into units suitable for 
our purpose by \citet{maiolino07}. We have then converted to AGN bolometric
luminosity adopting $L_{Bol}$=7$\nu L_\nu$(5100\AA) 
\citep[e.g.,][]{netzertr07}.
The star forming luminosities are based on the stacked greybody luminosities
for an adopted T=35K, $\beta$=1.5 SED at the individual redshift of each
source. For comparison, we add samples of low and high redshift QSOs 
\citep{netzer07,lutz08,page04,stevens05,priddey03,omont03}. The local universe ratio of black hole to bulge mass ratio of 0.14\% \citep{haering04} can 
be converted into a
formal `steady growth luminosity ratio' for star formation and AGN
$L_{SF}$/$L_{AGN}$=4.7$(0.1/\eta)$ that is shown
in Fig.~\ref{fig:growthplot} for a black hole accretion efficiency $\eta$=0.1. 

The points shown for the CDFS X-ray AGN reflect the stack averages, i.e., the
individual objects may scatter noticeably towards higher and lower star 
formation rates. Nevertheless, their location around the `continuous growth'
line is compatible with the picture of secular evolution outlined in the
previous section.
AGN accretion rates or star formation rates may here fluctuate to some extent
with time, moving individual sources around this line in 
Fig.~\ref{fig:growthplot}
in left/right and up/down direction, respectively, but
the location of the population overall does not require to place them 
in any special evolutionary state that would be deviating from a long 
term growth. Only the highest luminosity 
$L_{2-10keV}>10^{44}\ergs$ CDFS AGN approach the location of the
local and high redshift QSO in the comparison samples. Compared to the 
continuous growth ratio, 
these most luminous X-ray AGN as well as the QSOs are growing their black 
holes at a much faster rate, indicating that 
the matching star formation is most likely spread over longer timescales.
Again, we find that only the most luminous CDFS AGN match the possibly merger
related evolutionary pattern of QSOs.

The combination in Fig~\ref{fig:growthplot} of the CDFS AGN with the more 
luminous high redshift AGN observed in the (sub)mm by \citet{priddey03},
\citet{omont03}, \citet{page04} and \citet{stevens05} confirms the behaviour
discussed in Sect. 2.1. Star formation in the hosts of modest 
luminosity high-z AGN seems to depend little on the exact AGN luminosity -
reflecting the  'secular path'.
These external samples, however, extend the increase in star formation
above $L_{2-10keV}\sim 10^{44}$\ergs ($L_{AGN}\sim 3\times10^{45}$\ergs) 
that was already indicated in the CDFS data - reflecting the connection
between star formation and AGN on the `evolutionary connection/merger path'. 
The combined results for these samples and the CDFS AGN can be approximated
by a simple relation $L_{SF} = 10^{44.56}+10^{27.7}\times L_{AGN}^{0.38}$ 
(dashed line in Fig.~\ref{fig:growthplot}). 
Here, the smaller slope for the power law at 
high AGN luminosities compared to the slope 1 implied by the `steady growth 
ratio' plausibly reflects that at the highest AGN luminosities the AGN growth 
(at the time of observation) is increasingly faster compared to the host 
growth. While this simple two component --
constant plus power law -- parametrization is a plausible reflection of 
the two growth modes, it should be considered illustrative and
detailed parameters viewed with caution, given the 
sample selections and the fact that we here compare z$\sim$1 AGN at 
moderate luminosities and z$\sim$2 AGN at the highest luminosities. 

\subsection{Comparison to the local AGN population}

The two evolutionary paths outlined above for high redshift AGN from deep 
X-ray fields imply a straightforward consistency check with the local AGN 
population. If the hosts of moderate luminosity AGN at z$\sim$1 have star 
formation  rates similar to nonactive massive galaxies at the same redshift, 
the star formation rates in the host of moderate luminosity AGN should 
follow the decrease of star formation on the general massive galaxy 
population towards redshift zero.

To perform this check in a methodology as consistent with the LABOCA analysis
of (E)CDFS X-ray AGN as possible, we have used a local unbiased 14-150keV 
extremely hard X-ray selected AGN sample from the 39 month Palermo 
\swift\/-BAT catalog \citep[PSB,][see also 
Tueller et al. (2008) for an earlier BAT AGN catalog]{cusumano09} in 
conjunction with the \iras\ all-sky far-infrared survey. We selected 
(Shao et al., in preparation) from the
PSB survey sources classified as Seyferts, LINERs, quasars, and other AGN, 
explicitly omitting blazars. We excluded remaining objects with the 
possibility of a strong nonthermal contribution to the far-infrared on the 
basis of the NED SED, objects at galactic latitude $|b|<15$ and objects at 
redshift z$>$0.3 for which the \iras\ 60$\mu$m band no longer probes the 
rest-frame far-infrared. For the remaining 293 AGN we used  the 
\iras\ Faint Source Catalog 60$\mu$m detections where available, otherwise 
we used Scanpi\footnote{http://scanpi.ipac.caltech.edu:9000/applications/Scanpi/index.html} 
to obtain 60$\mu$m measurements for faint or individually nondetected objects.
We calculated rest frame 2--10keV luminosities extrapolating from the BAT 
fluxes and the redshift, assuming an AGN photon index of 1.8, and infrared 
luminosities $\nu L_\nu$(60$\mu$m) in the observed frame. Because about 20\%
of the sample are individually undetected at 60$\mu$m and for consistency with
the LABOCA stacking, we stacked the luminosities in seven 2--10keV luminosity 
bins spanning the PSB luminosity range with sufficient statistics in each bin.

Star forming and AGN luminosities for these local AGN stacks are shown in 
Fig.~\ref{fig:growthplot} in direct comparison to the (E)CDFS results.
At moderate AGN luminosities ($L_{AGN}< 3\times10^{44}$\ergs), the star 
forming luminosities of local hosts are about an order of magnitude lower 
than in z$\sim$1 (E)CDFS AGN hosts. On the other hand, these local AGN appear
to follow down to lower AGN luminosities the diagonal correlation of star 
forming and AGN luminosity that is reflecting the `merger path'.
This is in full agreement with previous local work, for example the 
infrared-based results of \citet{roro95} and \citet{netzer07} covering
the regime of local QSOs, and modelling based on SDSS
spectroscopy by \citet{netzer09}. The latter traces a 
correlation down to even lower AGN luminosities $\lesssim 10^{43}$\ergs\ where
star formation rates in our BAT stacks flatten out. We speculate this might
relate to a more difficult disentanglement of weak AGN and moderately
star forming galaxies galaxies in optical spectra than in very hard-X vs. 
far-infrared.   

\citet{mullaney10} use \spitzer\ 70$\mu$m data to trace the evolution of
the far-infrared to X-ray luminosity ratio of AGN to redshift z$\sim$2. 
They find no significant change of IR luminosity with redshift for 
luminous $L_{2-10keV}=10^{43-44}\ergs$ X-ray AGN but an 
increase for more modest  $L_{2-10keV}=10^{42-43}\ergs$ AGN. While an 
observed wavelength of 70$\mu$m makes for a difficult AGN/host diagnostic
at the high-z/high-l end of that range, where AGN heated dust will enter 
strongly the observed band, these results agree with our submm-based 
finding that an increase of host SFR from local towards z$\sim$1 occurs 
for modest luminosity AGN only. We note that the FIR luminosity increase by 
about an order of magnitude to z$\sim$1 would be difficult to reconcile 
with the 
alternative explanation of an AGN covering factor increase \citep{mullaney10},
since covering factors $<$0.1 would be required for local AGN.
Another recent study \citep{trichas09} addresses the starburst/AGN 
connection in luminous z$\sim$1 SWIRE AGN. Being based on rare 70$\mu$m
detections only it is not directly comparable to our work but consistent
with the presence of elevated star formation in 
$L_{2-10keV}\sim 10^{43}\ergs$ z$\sim$ AGN.

Our finding of elevated star formation in modest luminosity high z AGN
is consistent with the change of typical
star formation rates from z$\sim$1 to z=0 in the non-active massive galaxy
population \citep{noeske07,elbaz07,daddi07}. In Fig.~\ref{fig:growthplot},
both local and z$\sim$1 AGN are consistent with a `secular' path 
with star forming luminosity independent of AGN luminosity, and an 
`evolutionary connection' path with increasing star forming luminosity at 
high AGN luminosity. The evolutionary path, possibly linked to merging,
appears to emerge locally at lower AGN luminosities from the locally 
lower level of general star formation.

We have previously discussed the location of local and high redshift 
X-ray selected AGN and optical QSOs in the L$_{SF}$ vs. L$_{AGN}$
diagram of Fig.~\ref{fig:growthplot}. In the merger evolutionary picture
and in a sequence where the strongest star formation occurs 
before strongest AGN activity, galaxies should follow a loop path moving up  
in the L$_{SF}$ vs. L$_{AGN}$ diagram with the rise of star formation and 
then to the right and perhaps down towards the most intense AGN phase (see 
discussion in \citet{netzer09}). We have placed SMGs {\em with weak X-ray AGN}
 from \citet{alexander05a} in Fig.~\ref{fig:growthplot} to show that their 
location in comparison to QSOs is consistent with such a path. SMGs without 
detected AGN would be found yet further to the left in 
Fig.~\ref{fig:growthplot}. Comparing the location of individual submm-selected
SMGs hosting weak AGN to the X-ray selected CDFS stacks has to be done with 
caution since the CDFS stacks themselves 
will contain similar SMGs contributing to the upturn in mm emission at high 
X-ray luminosity. A thorough analysis of possible evolutionary loops in such
a L$_{SF}$ vs. L$_{AGN}$ diagram will thus have to be done on the basis of 
future individual measurements for sources in all the relevant parts of the 
diagram.  

\section{Conclusions}
We have used the combination of the LESS 870$\mu$m survey and deep X-ray 
surveys of the (E)CDFS region to study star formation in the hosts of AGN
covering a wide range of redshifts and luminosities centered around z$\sim$1
and $L_{2-10keV}\sim 10^{43}\ergs$. Stacking LESS data at the positions of 
all 895 AGN we detect at high
significance submm emission at S$_{870\mu m}$=0.49$\pm0.04$mJy 
corresponding to 
average star formation rates of about 30\Msun\,yr$^{-1}$. Using the good 
statistics to break down the sample according to AGN properties, we find an
increase with redshift and little change with AGN luminosity, except for 
indications for an upturn in host star formation at $L_{2-10keV}>10^{44}\ergs$.
The bulk of the X-ray AGN do not show changes with AGN obscuration as 
expected from the merger evolutionary scenario, but such a behaviour may 
emerge at $L_{2-10keV}\gtrsim 10^{44}\ergs$. 

Combined with results for higher luminosity AGN not properly sampled in 
the 0.25 square degree ECDFS, we conclude that the bulk of deep survey 
X-ray AGN seem to be hosted by galaxies evolving secularly, with star 
formation 
rates similar to comparably massive non-active galaxies and no close link
between AGN and global host star formation. In contrast, the most luminous    
$L_{2-10keV}>10^{44}\ergs$ AGN seem to follow a path where AGN activity and 
obscuration appear to be more closely linked to host star formation, likely via
merger evolution.

The properties of local \swift\/-BAT selected AGN with \iras\/-based 
far-infrared star forming luminosities are consistent with these two paths.
The host star formation rates of moderate luminosity AGN are decreasing from 
z$\sim$1 to z=0 similar to the decrease of SFR in non-active massive galaxies 
over this redshift interval.

\acknowledgements
We acknowledge helpful comments by an anonymous referee.
We thank Hagai Netzer and Sylvain Veilleux for discussions. D.L. thanks the 
Aspen Physics Center for hospitality during part of the 
preparation of this paper. I.R.S., K.E.K.C. R.J.I and J.W. acknowledge support
from STFC.
W.N.B, D.A.R. and Y.Q.C acknowledge support from CXC grant SP8-9003A.
Based on data obtained with the APEX telescope, with programme IDs
078.F-9028(A), 079.F-9500(A), 080.A-3023(A) and 081.F-9500(A).


%
%

\clearpage

\begin{deluxetable}{lllllcl}
\tablecolumns{7}
\tablewidth{0pt} 
\tablecaption{Stacking results for combined CDFS+ECDFS list}
\tablehead{
\colhead{Group}                &
\colhead{N$_{\rm Source}$}     &
\colhead{S$_{870\mu m}$}       &
\colhead{$\sigma_{map}$}       &
\colhead{$\sigma_{subsample}$ }&
\colhead{S$_{870\mu m,residual}$}&
\colhead{L$_{\rm IR}$}         \\
\colhead{}                 &
\colhead{}                 &
\colhead{mJy}              &
\colhead{mJy}              &
\colhead{mJy}              &
\colhead{mJy}              &
\colhead{10$^{11}$L$_\sun$}\\
\colhead{(1)}&\colhead{(2)}&\colhead{(3)}&\colhead{(4)}&
\colhead{(5)}&\colhead{(6)}&\colhead{(7)}\\
}
\startdata
\sidehead{No grouping}
All CDFS \& ECDFS          & 895&0.490&0.044&0.054&0.344&\\
\sidehead{Grouping by redshift}
z$\leq$1.2                 & 383&0.279&0.066&0.083&0.218&1.15\\
z$>$1.2                    & 365&0.637&0.068&0.085&0.445&3.97\\
z$>$2                      & 158&0.597&0.106&0.129&0.430&3.79\\
All z                      & 748&0.453&0.047&0.059&0.328&2.48\\
No good phot/spec z        & 147&0.695&0.108&0.134&0.434&\\ 
\tablecomments{
Col. (1) --- Group of X-ray AGN entering the stack.\\
Col. (2) --- Number of sources in stack.\\
Col. (3) --- Weighted mean of 870$\mu$m flux densities for stack. Measurements 
for stacks which are at $<3\sigma$, hence not significantly detected, are 
highlighted by italics.\\
Col. (4) --- Standard deviation of the weighted means, obtained by drawing 
many samples of N$_{Source}$ at random positions from the LABOCA map.\\
Col. (5) --- Standard deviation of the weighted means from drawing many subsamples of N$_{Source}$ from the full N=895 AGN sample.\\
Col. (6) --- Weighted mean of 870$\mu$m flux densities for stack on the 
residual map, obtained by subtracting all individual detections from the
original map. For the stacks from the residual map, the corresponding 
errors $\sigma_{map}$ and $\sigma_{subsample}$ (not shown) are consistent 
with each other and $\sim$12\% less than $\sigma_{map}$ for the full map.\\
Col. (7) --- Stacked IR luminosity, from the weighted mean of the luminosities
of rest frame T=35K $\beta$=1.5 greybodies individually matched to the 
redshift and 870$\mu$m flux of each X-ray AGN. Available only for samples with
redshifts for each object.
\\
}
\enddata
\label{tab:stackcombo}
\end{deluxetable}

\clearpage

\begin{deluxetable}{lllllcl}
\tablecolumns{7}
\tablewidth{0pt} 
\tablecaption{Stacking results for Tozzi et al. CDFS list}
\tablehead{
\colhead{Group}                &
\colhead{N$_{\rm Source}$}     &
\colhead{S$_{870\mu m}$}       &
\colhead{$\sigma_{map}$}       &
\colhead{$\sigma_{subsample}$} &
\colhead{S$_{870\mu m,residual}$}&
\colhead{L$_{\rm IR}$}         \\
\colhead{}                 &
\colhead{}                 &
\colhead{mJy}              &
\colhead{mJy}              &
\colhead{mJy}              &
\colhead{mJy}              &
\colhead{10$^{11}$L$_\sun$}\\
\colhead{(1)}&\colhead{(2)}&\colhead{(3)}&\colhead{(4)}&
\colhead{(5)}&\colhead{(6)}&\colhead{(7)}\\
}
\startdata
\sidehead{No grouping}
All CDFS                   & 302&0.437&0.074&0.094&0.328&2.32\\
\sidehead{Grouping by redshift}
z$\leq$1                   & 145&0.362&0.109&0.134&0.258&1.47\\
z$>$1                      & 157&0.506&0.104&0.128&0.394&3.13\\
z$>$2                      &  63&0.779&0.162&0.202&0.561&4.93\\
\sidehead{Grouping by intrinsic rest frame 2-10keV luminosity}
L$<10^{42}$erg s$^{-1}$    &  48&{\it 0.379}&0.189&0.233&0.289&1.05\\
L$<10^{43}$erg s$^{-1}$    & 138&0.361      &0.111&0.138&0.305&1.52\\
L$\geq 10^{43}$erg s$^{-1}$& 164&0.500      &0.102&0.126&0.348&3.01\\
L$<10^{44}$erg s$^{-1}$    & 260&0.341      &0.086&0.101&0.266&1.71\\
L$\geq 10^{44}$erg s$^{-1}$&  42&1.036      &0.200&0.250&0.718&6.24\\
\sidehead{Grouping by X-ray obscuring column from spectral fit}
N$_{H}< 10^{22}$cm$^{-2}$    & 114&{\it 0.315}&0.122&0.152&0.280&1.32\\
N$_{H}\geq 10^{22}$cm$^{-2}$ & 188&0.511      &0.095&0.120&0.357&2.94\\
N$_{H}<3\times 10^{22}$cm$^{-2}$   &169&0.421 &0.098&0.115&0.325&2.04\\
N$_{H}\geq 3\time 10^{22}$cm$^{-2}$&133&0.458 &0.113&0.140&0.333&2.66\\
N$_{H}<10^{23}$cm$^{-2}$     & 225&0.416      &0.087&0.108&0.310&2.08\\
N$_{H}\geq 10^{23}$cm$^{-2}$ &  77&0.498      &0.149&0.123&0.383&3.01\\
\sidehead{\parbox{10.cm}{Sources with $L_{2-10keV}\geq 3\ 10^{43}\ergs$, 
grouping by X-ray column}}
N$_{H}< 10^{22}$cm$^{-2}$   & 18&{\it 0.072}&0.316&0.378&0.185&0.40\\
N$_{H}\geq 10^{22}$cm$^{-2}$& 80&0.641      &0.147&0.180&0.427&3.95\\
N$_{H}< 10^{23}$cm$^{-2}$   & 48&{\it 0.400}&0.189&0.233&0.341&2.42\\
N$_{H}\geq 10^{23}$cm$^{-2}$& 50&0.667      &0.184&0.226&0.422&4.12\\
\sidehead{\parbox{10.cm}{Sources with $L_{2-10keV}\geq 1\ 10^{44}\ergs$ 
and spectroscopic redshift, grouping by X-ray column}}
N$_{H}< 10^{22}$cm$^{-2}   $&  6&{\it 0.433}&0.531&0.653&0.549&2.23\\
N$_{H}\geq 10^{22}$cm$^{-2}$& 21&1.390      &0.285&0.352&0.858&8.43\\
N$_{H}< 10^{23}$cm$^{-2}   $& 13&1.098      &0.366&0.447&0.860&6.40\\
N$_{H}\geq 10^{23}$cm$^{-2}$& 14&1.244      &0.346&0.432&0.719&7.57\\
\sidehead{\parbox{10.cm}{Sources with $L_{2-10keV}< 3\times 10^{43}\ergs$ and 
spectroscopic redshift, grouping by X-ray column}}
N$_{H}< 10^{22}$cm$^{-2}   $& 96&{\it 0.360}&0.133&0.166&0.298&1.47\\
N$_{H}\geq 10^{22}$cm$^{-2}$&108&0.415      &0.124&0.156&0.307&2.23\\
\sidehead{\parbox{10.cm}{Sources with $L_{0.5-2keV}$ or 
$L_{2-10keV}\geq 1\ 10^{44}\ergs$ and spectroscopic redshift, grouping 
by optical spectral type}}
Type 1 (BLAGN)              & 10&{\it 0.737}&0.410&0.508&0.743&4.29\\
Type 2                      & 17&      1.436&0.315&0.388&0.815&8.64\\
\tablecomments{Definition of columns as in  Table~\ref{tab:stackcombo}.
 Stacks measured at $<3\sigma$, hence not significantly detected, are 
highlighted by italics.}
\enddata
\label{tab:stacktozzi}
\end{deluxetable}
%
%


\begin{thebibliography}{999}
\bibitem[Alexander et al.(2001)]{alexander01} Alexander, D.M., Brandt, W.N.,
Hornschemeier, A.E., Garmire, G.P., Schneider, D.P., Bauer, F.E., 
Griffiths, R.E. 2001, \aj, 122, 2156
\bibitem[Alexander et al.(2003)]{alexander03} Alexander, D.M., et al.
2003, \aj, 125, 383
\bibitem[Alexander et al.(2005a)]{alexander05a} Alexander, D.M., Smail, I.,
Bauer, F.E, Chapman, S.C., Blain, A.W., Brandt, W.N., Ivison, R.J. 2005a,
\nat, 434, 738 
\bibitem[Alexander et al.(2005b)]{alexander05b} Alexander, D.M., Bauer, F.E., 
Chapman, S.C., Smail, I., Blain, A.W., Brandt, W.N., Ivison, R.J. 2005b,
\apj, 632, 736 
\bibitem[alexander et al.(2008)]{alexander08} Alexander, D.M., et al. 2008,
\aj, 135, 1958
\bibitem[Almaini et al.(2003)]{almaini03} Almaini, O., et al. 2003,
\mnras, 338, 303
\bibitem[Alonso-Herrero et al.(2008)]{alonso08} Alonso-Herrero, A., 
P\'erez-Gonz\'alez, P.G., Rieke, G.H., Alexander, D.M., Rigby, J.R.,
Papovich, C., Donley, J.L., Rigopoulou, D. 2008, \apj, 677, 127
\bibitem[Barger et al.(2001)]{barger01} Barger, A.J., Cowie, L.L.,
Steffen, A.T., Hornschemeier, A.E., Brandt, W.N., Garmire, G.P.
2001, \apj, 560, l23
\bibitem[Barger et al.(2003)]{barger03} Barger, A. et al. 2003, \aj, 126, 623 
\bibitem[Bauer et al.(2004)]{bauer04} Bauer, F.E., et al. 2004, \aj, 128 ,2048
\bibitem[Bavouzet(2009)]{bavouzet09} Bavouzet, N., 2009, Th\`ese de Doctorat,
Universit\'e Paris Sud 11
\bibitem[Blain et al.(2002)]{blain02} Blain, A.W., Smail, I., Ivison, R.J.,
Kneib, J.-P., Frayer, D.T. 2002, \physrep, 369, 111 
\bibitem[Borys et al.(2005)]{borys05} Borys, C., Smail, I., Chapman, S.C.,
Blain, A.W., Alexander, D.M., Ivison, R.J. 2005, \apj, 635, 853 
\bibitem[Bundy et al.(2008)]{bundy08} Bundy, K., et al. 2008, \apj, 681, 931
\bibitem[Chabrier(2003)]{chabrier03} Chabrier, G. 2003, \pasp, 115, 763
\bibitem[Colbert et al.(2002)]{colbert02} Colbert, E.J.M., Weaver, K.A., 
Krolik, J.H., Mulchaey, J.S., Mushotzsky, R.F. 2002, \apj, 202, 581
\bibitem[Cooper et al.(2008)]{cooper08} Cooper M.C., et al. 
2008, \mnras, 383, 1058
\bibitem[Coppin et al.(2005)]{coppin05} Coppin, K., Halpern, M., Scott, D.,
Borys, C., Chapman, S. 2005, \mnras, 357, 1022 
\bibitem[Coppin et al.(2006)]{coppin06} Coppin, K., et al. 2006, \mnras, 372,
1621  
\bibitem[Coppin et al.(2009)]{coppin09} Coppin, K., et al. 2009, \mnras, in 
press (arXiv 0902.4462)
\bibitem[Cusumano et al.(2009)]{cusumano09} Cusumano, G., 2009, AIP conference
proceedings 1126, 104
\bibitem[Daddi et al.(2007)]{daddi07} Daddi, E., et al. 2007, \apj, 670, 156
\bibitem[Elbaz et al.(2007)]{elbaz07} Elbaz, D., et al. 2007 \aap, 468, 33
\bibitem[Elvis et al.(1994)]{elvis94} Elvis, M., et al. 1994, \apjs, 95, 1 
\bibitem[Evans et al.(2001)]{evans01} Evans, A.S, Frayer, D.T., Surace, J.A.,
Sanders, D.B. 2001, \aj, 121, 3286
\bibitem[Fabian(1999)]{fabian99} Fabian, A.C. 1999, MNRAS, 308, L39
\bibitem[Fabian et al.(2000)]{fabian00} Fabian, A.C., et al. 2000, \mnras,
315, L8
\bibitem[F\"orster Schreiber et al.(2006)]{foerster06} F\"orster Schreiber, 
N.M., et al. 2006, \apj, 645, 1062
\bibitem[Genzel et al.(2008)]{genzel08} Genzel, R., et al. 2008, \apj,
687, 59 
\bibitem[Giacconi et al.(2002)]{giacconi02} Giacconi, R., et al. 2002, 
\apjs, 139, 369
\bibitem[Granato et al.(2004)]{granato04} Granato, G.L., de Zotti, G.,
Silva, L., Bressan, A., Danese, L. 2004, \apj, 600, 580
\bibitem[Greve et al.(2009)]{greve09} Greve, T., et al. 2009, \apj, submitted
(arXiv:0904.0028)
\bibitem[Grogin et al.(2005)]{grogin05} Grogin, N.A., et al. 2005, \apj,
627, L97 
\bibitem[G\"usten et al.(2006)]{guesten06} G\"usten, R., Nyman, L.\AA., 
Schilke, P., Menten, K., Cesarsky, C., Booth, R. 2006, \aap, 454, L13
\bibitem[H\"aring \& Rix(2004)]{haering04} H\"aring, N., Rix, H.W. 2004,
\apj, 604, L89 
\bibitem[Hasinger(2008)]{hasinger08} Hasinger, G., 2008, \aap, 490, 905
\bibitem[Holland et al.(1999)]{holland99} Holland, W.S., et al. 1999, \mnras,
303, 659
\bibitem[Hopkins et al.(2006)]{hopkins06} Hopkins, P.F., Hernquist, L.,
Cox, T.J., Di Matteo, T., Robertson, B., Springel, V. 2006, \apjs, 163, 1
\bibitem[Hopkins \& Hernquist(2009)]{hopkins09} Hopkins, P.F., Hernquist, L. 
2009, \apj, 694, 599
\bibitem[Hornschemeier et al.(2000)]{hornschemeier00} Hornschemeier, A.E., 
et al. 2000, \apj, 541, 49
\bibitem[Houck et al.(2005)]{houck05} Houck, J., et al. 2005, \apj, 622, L105
\bibitem[Ivison et al.(2007)]{ivison07} Ivison, R.J., et al. 2007, \mnras,
380, 199
\bibitem[Kauffmann et al.(2003)]{kauffmann03} Kauffmann, G., et al. 2003,
\mnras, 346, 1055 
\bibitem[Kennicutt(1998)]{kennicutt98} Kennicutt, R.C. 1998, \araa, 36, 189 
\bibitem[Koekemoer et al.(2004)]{koekemoer04} Koekemoer, A., et al. 2004,
\apj, 600, L123
\bibitem[Lehmer et al.(2005)]{lehmer05} Lehmer, B.D., et al. 2005, \apjs,
161, 21
\bibitem[Lehmer et al.(2008)]{lehmer08} Lehmer, B.D., et al. 2008, \apj,
681, 1163
\bibitem[Le Fevre et al.(2004)]{lefevre04} Le Fevre, O., et al. 2004, \aap,
428, 1043
\bibitem[Le Floc'h et al.(2001)]{lefloch01} Le Floc'h, E., Mirabel, I.F.,
Laurent, O., Charmandaris, V., Gallais, P., Sauvage, M., Vigroux, L.,
Cesarsky, C. 2001, \aap, 367, 487
\bibitem[Luo et al.(2008)]{luo08} Luo, B., et al. 2008, \apjs, 179, 19 
\bibitem[Lutz et al.(2005)]{lutz05} Lutz, D., et al. 2005, \apj, 625, L83
\bibitem[Lutz et al.(2008)]{lutz08} Lutz, D., et al. 2008, \apj, 684, 853
\bibitem[Maiolino et al.(2007)]{maiolino07} Maiolino, R., et al. 2007,
\aap, 468, 979
\bibitem[Mainieri et al.(2005)]{mainieri05} Mainieri, V., et al. 2005, \mnras,
356, 1571
\bibitem[Mainieri et al.(2005a)]{mainieri05a} Mainieri, V., et al. 2005a, \aap,
437, 805
\bibitem[Marconi \& Hunt(2003)]{marconi03} Marconi, A., Hunt, L.K, 2003,
\apj, 689, L21
\bibitem[Miller et al.(2008)]{miller08} Miller, N., Fomalont, E.B., 
Kellermann, K.I., Mainieri, V., Norman, C., Padovani, P., Rosati, P.,
Tozzi, P. 2008, \apjs, 179, 114 
\bibitem[Mullaney et al.(2010)]{mullaney10} Mullaney, J.R., Alexander, D.M.,
Huynh, M., Goulding A.D., Frayer, D. 2010, \mnras, 401, 995
\bibitem[Nandra et al.(2007)]{nandra07} Nandra, K., et al. 2007, \apj, 
660, L11 
\bibitem[Negrello et al.(2005)]{negrello05} Negrello, M., Gonz\'alez-Nuevo, J.,
 Magliochetti, M., Moscardini, L., de Zotti, G., Toffolatti, L., Danese, L.
2005, \mnras, 358, 869 
\bibitem[Nenkova et al.(2008)]{nenkova08} Nenkova, M., Sirocky, M.M., 
Nikutta, R., Ivezic, Z., Elitzur, M. 2008, \apj, 685, 160
\bibitem[Netzer \& Trakhtenbrot(2007)]{netzertr07} Netzer, H., Trakhtenbrot, B.
2007 \apj, 564, 754
\bibitem[Netzer et al.(2007)]{netzer07} Netzer, H., et al. 2007, \apj, 666, 806
\bibitem[Netzer(2009)]{netzer09} Netzer, H. 2009, \mnras, 399, 1907
\bibitem[Noeske et al.(2007)]{noeske07} Noeske, K.G., et al. 2007, \apj, 660, 
L43
\bibitem[Norman et al.(2002)]{norman02} Norman, C., et al. 2002, \apj, 571,218
\bibitem[Omont et al.(2003)]{omont03} Omont, A., Beelen, A., Bertoldi, F.,
Cox, P., Carilli, C.L., Priddey, R.S., McMahon, R.G., Isaak, K.G. 2003,
\aap, 398, 857
\bibitem[Page et al.(2001)]{page01} Page, M.J., Stevens, J.A., 
Mittaz, J.P.D., Carrera, F.J. 2001, Science 294, 2516
\bibitem[Page et al.(2001b)]{page01b} Page, M.J., Mittaz, J.P.D., Carrera, F.J.
2001, \mnras, 325, 575
\bibitem[Page et al.(2004)]{page04} Page, M.J., Stevens, J.A., Ivison, R.J.,
Carrera, F.J. 2004, \apj, 611, L85 
\bibitem[Papadopoulos \& Seaquist(1999)]{papadopoulos99} Papadopoulos, P.P.,
Seaquist, E.R. 1999, \apj, 514, L95
\bibitem[Pier \& Krolik(1992)]{pier92} Pier, E.A., Krolik, J.H. 1992, 
\apj, 401, 99
\bibitem[Pier \& Krolik(1993)]{pier93} Pier, E.A., \& Krolik, J.H.\ 1993, 
\apj, 418, 673
\bibitem[Pierce et al.(2007)]{pierce07} Pierce, C.M., et al. 2007, \apj
660, L19 
\bibitem[Priddey et al.(2003)]{priddey03} Priddey, R.S., Isaak, K.G.,
McMahon, R.G., Omont, A. 2003, \mnras, 339, 1183
\bibitem[Rigopoulou et al.(2009)]{rigopoulou09} Rigopoulou, D., et al. 2009,
\mnras, 400, 1199
\bibitem[Roche et al.(2006)]{roche06} Roche, N.D., Dunlop, J., Caputi, K.L.,
McLure, R., Willott, C.J., Crampton, D. 2006, \mnras, 370, 74
\bibitem[Rovilos et al.(2007)]{rovilos07} Rovilos, E., Georgakakis, A.,
Georgantopoulos, I., Afonso, j., Koekemoer, A.M., Mobasher, B., Goudis, C.
2001, \aap, 466, 119
\bibitem[Rowan-Robinson(1995)]{roro95} Rowan-Robinson, M., 1995, \mnras,
272, 737
\bibitem[Sanders et al.(1988)]{sanders88} Sanders, D.B., Soifer, B.T.,
Elias, J.H., Madore, B.F., Matthews, K., Neugebauer, G., Scoville, N.Z.
1988, \apj, 325, 74
\bibitem[Schweitzer et al.(2006)]{schweitzer06} Schweitzer, M., et al. 
2006, \apj, 649, 79
\bibitem[Severgnini et al.(2000)]{severgnini00} Severgnini, P., et al.
2000, \aap, 360, 457
\bibitem[Shapiro et al.(2008)]{shapiro08} Shapiro, K.L, et al. 2008, \apj, 
682, 231 
\bibitem[Silverman et al.(2008)]{silverman08} Silverman, J.D., et al. 2008,
\apj, 675, 1025
\bibitem[Silverman et al.(2009)]{silverman09} Silverman, J.d., et al. 2009,
\apj, 696, 396
\bibitem[Siringo et al.(2009)]{siringo09} Siringo, G., et al. 
2009, \aap, 497, 945
\bibitem[Solomon \& VandenBout(2005)]{solomon05} Solomon, P.M., VandenBout, 
P.A. 2005, \araa, 43, 677
\bibitem[Steffen et al.(2006)]{steffen06} Steffen, A., et al. 2006, \aj,
131, 2826 
\bibitem[Stevens et al.(2005)]{stevens05} Stevens, J.A., Page, M.J.,
Ivison, R.J., Carrera, F.J., Mittaz, J.P.D., Smail, I., McHardy, I.M.
2005, \mnras, 360, 610
\bibitem[Sturm et al.(2006)]{sturm06} Sturm, E., Hasinger, G., Lehmann, I.,
Mainieri, V., Genzel, R., Lehnert, M.D., Lutz, D., Tacconi, L.J.  2006, 
\apj, 642, 81
\bibitem[Szokoly et al.(2004)]{szokoly04} Szokoly, G., et al. 2004,
\apjs, 155, 271
\bibitem[Tozzi et al.(2006)]{tozzi06} Tozzi, P., et al. 2006, \aap, 451, 457
\bibitem[Treister et al.(2009)]{treister09} Treister, E., et al. 2009, \apj,
691, 1713
\bibitem[Trichas et al.(2009)]{trichas09} Trichas, M., Georgakakis, A.,
Rowan-Robinson, M., Nandra, K., Clements, D., Vaccari, M. 2009, \mnras,
399, 663
\bibitem[Tueller et al.(2008)]{tueller08} Tueller, J., Mushotzky, R.F., 
Barthelmy, S., Cannizzo, J.K., Gehrels, N., Markwardt, C.N., Skinner, G.K.,
Winter, L.M., 2008, \apj, 681, 113 
\bibitem[Waskett et al.(2003)]{waskett03} Waskett, T.J., et al.
2003, \mnras, 341, 1217
\bibitem[Wei\ss\ et al.(2009)]{weiss09} Wei\ss\/, A., et al., 2009, \apj,
707, 1201
\bibitem[Wolf et al.(2004)]{wolf04} Wolf, C., et al. 2004, \aap, 421, 913
\bibitem[Wolf et al.(2008)]{wolf08} Wolf, C., Hildebrandt, H., Taylor, E.N.,
Meisenheimer, K., 2008, \aap, 491, 933 
\bibitem[Zakamska et al.(2008)]{zakamska08} Zakamska, N.L, G\'omez, L.,
Strauss, M.A., Krolik, J.H. 2008, \aj, 136, 1607
\end{thebibliography}
\end{document}